\documentclass[12pt]{iopart}
\usepackage{graphicx}
\usepackage{dcolumn}
\usepackage{bm}
\usepackage{iopams}
\usepackage{cite}
\usepackage{color}

\begin{document}
\title{Nucleon-Nucleus Optical Potential computed with the Gogny interaction}
\author{Juan Lopez Mora\~{n}a - Xavier Vi\~{n}as$^*$}
\address{Departament de F\'isica Qu\`antica i Astrof\'isica and Institut de Ci\`encies del Cosmos (ICCUB),
Facultat de F\'isica, Universitat de Barcelona, Mart\'i i Franqu\`es 1, E-08028 Barcelona, Spain}
$^*$E-mail: xavier@fqa.ub.edu (corresponding author)\\
\date{\today}
\setlength\mathindent{20pt}
\begin{abstract} \\ \\

The ability of the Gogny forces of the D1 family to describe the nucleon-nucleus scattering is studied. To this end, we use an optical model potential built up using a semi-microscopic nuclear matter approach. The real and imaginary parts of the optical model are provided by the first and second-order terms, respectively, of the Taylor expansion of the mass operator calculated within the Brueckner-Hartree-Fock method using the reaction G-matrix built up with the effective Gogny force instead of a microscopic interaction. The optical potential in finite nuclei is obtained through the Local Density Approximation using the neutron and proton densities provided by a quasi-local Hartree-Fock calculation with the same Gogny force for the sake of consistency. A reasonable good agreement is found between the theoretical differential cross-sections and the analyzing powers of the elastic neutron and proton scattering along the periodic table from Ca to Pb calculated with the Gogny forces and the corresponding values predicted by the global phenomenological potential of Koning and Delaroche. To investigate the limits of the approximations used in this work, comparisons with the results of nucleon-nucleus elastic scattering in $^{40}$Ca and  $^{48}$Ca obtained using the Nuclear Structure Model are also performed. 
\end{abstract}

\section{Introduction}

 From the early days of the nuclear physics nuclear reactions have been an important source of information about the structure and dynamics of the atomic nucleus. In this respect, the so-called Optical Model (OM) \cite{hodgson63} is one of the most important theoretical tools to deal with the analysis of nuclear scattering experiments induced by nucleons, deuterons, alpha particles and heavy ions. The OM is based on the simplified hypothesis that the individual nucleon-nucleon interactions between the projectile and the target can be replaced by a complex mean-field potential, whose real part describes the shape elastic scattering and the imaginary part takes into account the particles separated from the elastic channel by inelastic, pre-equilibrium and compound nucleus processes. The OM potentials can be classified into two wide groups, the phenomenological optical potentials (POP) and the microscopic optical potentials (MOP). In the POP one assumes some analytical form for the  radial dependence of the potential, usually of the Woods-Saxon type, and fits its parameters by minimizing the difference between the theoretical predictions and the experimental results of a selected set of measurements of scattering observables in some nuclear reactions. There are several ways of obtaining the parameters of the POP depending on fitting a single nucleus and a single energy (local models) or a set of nuclei in a range of masses and energies (global models). Well known examples of these global POP are the fits of Becchetti and Greenless \cite{becchetti69}, Varner et al \cite{varner91} and Koning and Delaroche \cite{koning03}. Usually the POP predict quite accurately experimental scattering data within the domain of mass and atomic numbers of the target and bombarding energy of the projectile where the parameters of the potential have been fitted. However, the predictions of these POP to another regions with different targets and/or energies of the incident nucleon may be doubtful. \\

A lot of work has been done in order to establish the OM from  more microscopic grounds. It started  sixty years ago when it was pointed out that the MOP could be identified with the mass operator of the one-particle Green function \cite{bell59}. Using this fact, Jeukene, Lejeune and Mahaux (JLM) obtained a MOP performing Brueckner-Hartree-Fock (BHF) calculations in nuclear matter with a realistic nucleon-nucleon (NN) interaction \cite{jeukene74,jeukene76,jeukene77a,jeukene77b}. In this approximation one first derives in nuclear matter the mass operator $M$, which is a function of the in-medium density of the nucleus and of the momentum $\vec{k}$ and energy $E$ of the projectile. Next, this MOP  is applied to finite nuclei through the Local Density Approximation (LDA). It is found that this model can be used in an energy domain of the incident nucleon up to 160 MeV within a mass number range $12\le A \le 208$. A possible improvement of this MOP to reproduce experimental data of nucleon scattering by finite nuclei has been proposed in Ref.\cite{bauge98}, where the authors add a spin-orbit contribution to the MOP of JLM and renormalize the real and imaginary parts of the MOP by suitable factors in order to obtain a good overall description of the neutron and proton scattering and reaction data up to 200 MeV in line with the predictions of POP. Very recently, nucleon-nucleus MOP based on the JLM scheme and the effective field theory  \cite{whitehead19,whitehead20} has been developed and applied to describe elastic proton and neutron scattering by Ca isotopes. Also {\it ab initio} optical potentials have been obtained using the self-consistent Green's function theory \cite{idini19} and the Green's function approach within the coupled-cluster method \cite{rotureau18}.\\

The use of effective nuclear forces of Skyrme type \cite{vautherin72} in the nuclear reactions scenario was considered first by Dover and Giai \cite{dover72}, who investigated the connection between the self-consistent mean-field in the Hartree-Fock (HF) approximation and the real part of the OM potential. Effective interactions, as for instance the Skyrme \cite{vautherin72} or Gogny \cite{decharge80} forces, which reproduce accurately the ground state properties of finite nuclei along the whole periodic table, can be regarded as effective parametrizations of the G-matrix \cite{decharge80}. Therefore, it seems reasonable to use such a kind of interaction to obtain the MOP in the nuclear matter (NM) approach through the calculation of the first and second order terms of the expansion of the mass operators, which can be associated to the real and imaginary parts of the MOP, respectively. Work in this direction has been developed in the recent years by two different groups. In Refs.\cite{shen81,shen09,pilipenko10}, it is shown that in the NM approach several Skyrme interactions are able to describe the nucleon-nucleus  elastic scattering in more reasonable agreement with the experiment without including any additional parameter. These findings have motivated  these groups to propose new Skyrme interactions, namely SKOP1-SKOP6 \cite{pilipenko10,pilipenko12} and SkC-SkD \cite{xu14}, which are  fitted to reproduce simultaneously experimental NM and finite nuclei properties as well as nucleon-nucleus elastic scattering data. In all these works the imaginary part of the central term of the MOP, which also includes a spin-orbit contribution, is derived in the NM approach and then applied to finite nuclei through the LDA. In Refs.\cite{shen09,xu14} the real part of the MOP is computed in two different ways. In one of them, the authors use strictly the NM approach and in the other one the real part is given by the Hartree-Fock (HF) mean-field of the target, which includes surface terms that does not appear in the pure NM approach. The real part of the spin-orbit potential, which is relevant in the case of nucleon-nucleus scattering, is also provided by the Skyrme-HF calculation in the target nucleus. In Refs. \cite{pilipenko10,pilipenko12} the real part of the MOP is again computed with the HF mean-field associated to the Skyrme interactions, but in this case the authors also take into account the rearrangement terms, which are not included in the real part of the MOP in Refs.\cite{shen81,shen09,xu14}. \\

Another different approach for deriving the MOP in finite nuclei is the so-called Nuclear Structure Model (NSM) introduced by N. Vinh Mau in Ref.\cite{vinhmau70}. In this approximation the OM potential consists of the sum of the HF mean-field and the polarization term. The HF potential is provided by the underlying NN interaction, and it is real, local, momentum-dependent and energy independent. The polarization potential, which is complex and energy dependent, appears due to intermediate excitations of the nucleus and describes the coupling of the entrance channel with inelastic channels. The NSM potential can be obtained using the self-consistent HF method and the random-phase approximation (RPA) \cite{blanchon15a,blanchon15b,blanchon17}. A different version of the NSM also includes the particle-vibration coupling on top of the RPA calculations \cite{hao15,hao18}. The NSM takes explicitly into account specific nuclear structure information of low-lying excited states and giant resonances in the target nucleus and may be relevant to describe the scattering at low energy. Due to the complexity of the calculation of the polarization term, there are few NSM calculations available in the literature. They have been basically applied to describe proton and neutron scattering in $^{40}$Ca and $^{48}$Ca using the Gogny D1S force \cite{blanchon15a,blanchon15b,blanchon17} and neutron scattering in $^{16}$O, $^{40}$Ca, $^{48}$Ca and $^{208}$Pb with the Skyrme SLy5 force \cite{hao15,hao18}. \\

Our aim in this work is to check the ability of the Gogny interactions without any additional parameter to give a global description of nucleon-nucleus scattering along the periodic table from Ca to Pb using the NM approach. To this end we use the D1S parametrization as the representative force. To obtain the MOP in a finite nucleus we use, on top of the NM matter approach, the LDA with the neutron and proton densities of the target nucleus. These densities are obtained by a quasilocal HF calculation \cite{gridnev98,soubbotin00,soubbotin03} with the same Gogny force (see Figure \ref{fig16} in Appendix 2). In this way the real and imaginary central parts of the MOP becomes functions of the radial coordinate $r$. We are aware that the pure LDA underestimates the surface diffuseness of the MOP \cite{jeukene77b}. To remedy this deficiency, a folding of the MOP with a Gaussian form-factor with a length scale that simulates the finite-range of the nuclear force has been proposed \cite{jeukene77b,bauge98}. This correction implies two additional parameters, namely the length scales used in the folding of the real and imaginary contributions to the MOP. However, we will not consider this correction in the present work because we are interested in investigating the ability of the Gogny forces, without any additional parameter, to describe the nucleon-nucleus scattering. The same HF calculation also provides the spin-orbit and Coulomb contributions to the real part of the MOP generated by the target nucleus (see Appendix 2). As in similar works at low energy using effective interactions \cite{shen81,shen09,xu14,pilipenko10,pilipenko12} or using chiral forces \cite{whitehead19,whitehead20}, the imaginary contribution to the spin-orbit term in the MOP has not been considered in our approach as far as it has little effect on the elastic cross-section \cite{brieva78}. In this study we perform two different types of comparisons. On the one hand, we compare our results with the predictions of the POP of Koning and Delaroche. On the other hand, we explore the limits of the predictive power of the NM approach using Gogny forces. To this end we compare our results with those obtained through the NSM in the case of neutron and proton scattering on $^{40}$Ca and $^{48}$Ca, for which detailed theoretical studies are available \cite{blanchon15a,blanchon15b,blanchon17}. The paper is organized as follows. We first discuss the theoretical framework, paying particular attention to the calculation of the real and imaginary parts of the central potential within the NM approach. Next we compare the differential cross sections and analyzing powers obtained with our model with the predictions of the POP of Koning and Delaroche. We also study as a function of the energy of the projectile the total and reaction cross sections and the volume integrals of the real and imaginary parts of the central contribution to the Gogny based MOP. The differential cross sections for nucleon scattering on $^{40}$Ca and $^{48}$Ca targets obtained with our model are compared with the more elaborated NSM results reported in Refs.\cite{blanchon15a,blanchon15b,blanchon17}. The summary and outlook are given in the last section.

\section{The microscopic optical potential in the nuclear matter approximation using Gogny forces}

We summarize now the basic theory used to derive the MOP in the NM approximation. Based on Ref.\cite{bell59} and following the JLM procedure, we start from the reaction $G$ matrix, which represents the interaction between two nucleons in the nuclear medium in a similar way that the $T$ matrix describes the scattering of two free nucleons. The $G$ matrix fulfills the so-called Bethe-Goldstone equation, which reads

\begin{equation}
G=V+V\frac{Q_p}{E_0-H_0}G \quad\rightarrow\quad G=V\left[1-\frac{Q_p}{E_0-H_0}V\right]^{-1}, \label{eq1}
\end{equation}

with $V$ the nucleon-nucleon interaction and $Q_P$ the Pauli operator, which is defined as 

\begin{equation}
Q_p=\sum_{\lambda,\mu > k_F}|\lambda,\mu><\lambda,\mu|, \label{eq2}
\end{equation}

where the sum in this expression excludes all the occupied states below the Fermi energy. As mentioned before, the essential quantity to obtain the MOP is the mass operator \cite{bell59}, which in the Brueckner-Hartree-Fock approximation can be written as \cite{jeukene76,jeukene77b}: 

\begin{equation}
M_{\alpha,\alpha} = \sum_{\rho}<\alpha\rho|G|\widetilde{\alpha\rho}>, \label{eq3}
\end{equation} 

where the antisymmetric matrix elements $<\alpha\rho|G|\widetilde{\alpha\rho}>$ are calculated in NM between plane wave states. Expanding now (\ref{eq3}) in Taylor series up to second order one obtains

\begin{equation}
M_{\alpha,\alpha} = M^{(1)}_{\alpha,\alpha} + M^{(2)}_{\alpha,\alpha} + \ldots\ldots \label{eq4}
\end{equation} 

The first and second terms of this expansion are given by

\begin{equation}
M^{(1)}_{\alpha,\alpha} = \sum_{\rho} n_{\rho}<\alpha\rho|V|\widetilde{\alpha\rho}>, \label{eq5}
\end{equation}

and

\begin{equation}
M^{(2)}_{\alpha,\alpha}=\frac{1}{2}\sum_{\nu,\lambda,\mu}<\alpha\nu|V|\widetilde{\lambda\mu}>\frac{(1 - n_{\lambda})(1 - n_{\mu})n_{\nu}}{\epsilon_{\alpha}+\epsilon_{\nu}-\epsilon_{\lambda}-\epsilon_{\mu}+i\eta}<\lambda\mu|V|\widetilde{\alpha\nu}>, \label{eq6}
\end{equation}

respectively. To derive (\ref{eq5}) and (\ref{eq6}) we have considered that the interaction $V$ is purely two-body and that the only relevant Feynman diagrams are the ones  depicted in Fig.1 of Ref.\cite{xu14}, where only two-body interactions have been considered. In this respect, let us mention that the second order diagrams contributing to Eq.(\ref{eq6}) correspond to uncorrelated 2p-1h states in the intermediate processes. Next, and following Refs. \cite{shen81,shen09,xu14}, we identify $M^{(1)}_{\alpha,\alpha}$, which coincides with the mean-field HF potential felt by the projectile, with the real part of the optical potential: 

\begin{equation}
V_{\alpha}=\sum_{\rho \leq k_F}<\alpha\rho|V|\widetilde{\alpha\rho}> \label{eq7},
\end{equation}

where the summation runs over all occupied states $\rho$ below the Fermi energy. The imaginary part of the optical potential is associated to the imaginary part of $M^{(2)}_{\alpha,\alpha}$, i.e.

\begin{equation}
W_{\alpha}=\frac{1}{2}Im \sum_{\nu\leq k_F \atop \lambda,\mu > k_F}<\alpha\nu|V|\widetilde{\lambda\mu}>\frac{1}{\epsilon_{\alpha}+\epsilon_{\nu}-\epsilon_{\lambda}-\epsilon_{\mu}+i\eta}<\lambda\mu|V|\widetilde{\alpha\nu}>. \label{eq8}
\end{equation}

In Eqs.(\ref{eq2})-(\ref{eq8}) Greek symbols correspond to the different states involved in the calculation, which are plane waves characterized by its momentum as well as by its spin and isospin, $\alpha$ being the state corresponding to the projectile. In these equations $n_{\rho}, n_{\lambda}, n_{\mu}$ and $n_{\nu}$ are the occupancies of the different states, which we assume to be 1(0) for states  below (above) the corresponding Fermi level. The tilde symbol indicates the antisymmetrized matrix element. In this equation  $\nu$ is an occupied state of the target that can be scattered to an unoccupied state $\mu$ above the Fermi level and $\lambda$ corresponds to the intermediate state of the projectile also above the Fermi level. \\

In our approach we obtain the MOP by using an effective Gogny force of the D1 family \cite{decharge80}. As we mentioned before, the Gogny interaction $V$ can be considered as a phenomenological realization of the $G$ matrix. In this sense, it is clear that the use of the Gogny force in the mean-field is fully justified and does not imply any double counting. This is also true for the imaginary part, where the Gogny force appears quadratically. It can be shown that the mass operator holds an analogy to the optical theorem, which relates the imaginary part of the free $T$ matrix to its absolute square value \cite{danielewicz94}. \\

Gogny forces were introduced by D. Gogny in the early eighties aimed to describe simultaneously the mean field and the pairing field with the same interaction. The Gogny D1S parametrization \cite{berger91} has been used in large-scale Hartree-Fock-Bogoliubov calculations of ground-state properties of finite nuclei along the whole periodic table \cite{cea}. A detailed analysis of these results shows some deficiencies in the theoretical description of the masses of neutron rich nuclei compared with the corresponding experimental values (see \cite{pillet17} for more details). To remedy these limitations of D1S, new parametrizations of the Gogny force, namely D1N \cite{chappert08}  and D1M \cite{goriely09}, have been proposed. These forces incorporate in their fitting protocol the constraint of reproducing, qualitatively, the microscopic Equation of State of Friedman and Pandharipande in neutron matter in order to improve the description of neutron rich nuclei. \\

The Gogny forces of the D1 family consists of a finite-range part and a zero-range density-dependent term together with a spin-orbit interaction, which is also zero-range  as in the case of Skyrme forces. The finite-range part is the sum of two Gaussian form factors with different ranges, each multiplied by all the possible spin-isospin exchange operators with different weights. This type of Gogny forces reads:

\begin{eqnarray}
&&V(\vec{r}_{12})=t_3(1+\hat{P}_\sigma)\delta(\vec{r}_{12})\left[\rho\left(\frac{\vec{r}_1+\vec{r}_2}{2}\right)\right]^{1/3} +\sum_{k=1}^{k=2}e^{-\left(\frac{\vec{r}_{12}}{\mu_k}\right)^2}\nonumber \\ &&\times\left(W_k+B_k\hat{P}_\sigma-H_k\hat{P}_\tau-M_k\hat{P}_\sigma\hat{P}_\tau\right)+iW_{LS}\left(\hat{\sigma}_1+\hat{\sigma}_2\right)\cdot\hat{k}^{\dag}\times\delta\left(\vec{r}_{12}\right)\hat{k},\label{eq9}
\end{eqnarray}

where
\[\hat{P}_\sigma=\frac{1}{2}\left(1+\hat{\sigma}_1\cdot\hat{\sigma}_2\right),\quad \textrm{and} \quad\hat{P}_\tau=\frac{1}{2}\left(1+\hat{\tau}_1\cdot\hat{\tau}_2\right),\]
are the spin and isospin exchange operators, respectively, while
\[\vec{r}_{12}=\vec{r}_1-\vec{r}_2,\quad \textrm{and} \quad\hat{k}=\frac{1}{2i}\left(\vec{\nabla}_1-\vec{\nabla}_2\right),\]

are the relative coordinate and the relative momentum of the two nucleons, respectively. The parameters of the force, namely $W_k, B_k, H_k, M_k, \mu_k$ ($k$ =1, 2), $t_3$, and $W_{LS}$, are fitted to reproduce some properties of finite nuclei and infinite nuclear matter (see Refs.\cite{decharge80} and \cite{goriely09} for more details about the fitting protocol of Gogny interactions). In Table 1 we give the parameters of the D1S interaction. 

\begin{table}
\begin{center}
\begin{tabular}{|l|l|l|l|l|l|l|} \hline & $k$ &$\mu_k(fm)$&$W_k$&$B_k$&$H_k$&$M_k[MeV]$ \\ \hline D1S &1 &0.7&-1720.3&1300&-1813.53&1397.60 \\ & 2&1.2 &103.64&-163.48&162.81&-223.93 \\ \hline
\end{tabular}
\end{center}

\begin{center}
\begin{tabular}{|l|l|l|}\hline D1S & $W_{LS}=130[MeV\cdot fm^5]$ & $t_3=1390[MeV\cdot fm^4]$  \\ \hline
\end{tabular}
\end{center}
\label{table1}
\caption{Parameters of the Gogny D1S force.}
\end{table}

Extrapolation of effective forces to domains different from the ones where the parameters of the interaction have been determined should be checked carefully. For instance, the extrapolation of Gogny interactions to the astrophysical domain fails because the most successful parametrization for describing finite nuclei, namely the D1N and D1M forces, are not able to predict the observed lower limit of the maximum mass of neutron stars of $2M_\odot$ \cite{gonzalez17}. However, the renormalization of the D1M Gogny interaction proposed in \cite{gonzalez18} allows to obtain a Gogny force suitable for astrophysical applications without losing its predictive power for finite nuclei. Thus it seems to be in order to check the ability of the Gogny forces to describe nucleon-nucleus reactions by exploring wide domains of mass and atomic numbers of the target and a broad range of energies of the projectile. This can provide additional information to the previous NSM results using the D1S Gogny force reported in Refs.\cite{blanchon15a,blanchon15b,blanchon17}.

\subsection{The real part of the optical potential}
Let us first discuss the NM approach. In this case a single-particle state, labeled with the index $\alpha$, is described by the following normalized wave function

\begin{equation}
\Psi_\alpha(\vec{r})=\frac{1}{\sqrt{\Omega}}e^{i\vec{k}\cdot\vec{r}}\chi_{\sigma\alpha}\zeta_{\tau\alpha};\qquad\sigma=\frac{1}{2},-\frac{1}{2}; \qquad\tau=\frac{1}{2}\;(n),-\frac{1}{2}\;(p) \label{eq10}
\end{equation}

where $\chi_{\sigma\tau}$ and $\zeta_{\sigma\tau}$ are the normalized spin and isospin spinors, respectively and $\Omega$ is the normalization volume. Notice that in phase space the state number is given by 

\begin{equation}
\sum_{\sigma\tau} \frac{\Omega \int d\vec{k}}{(2\pi)^3} \label{eq11}.
\end{equation}

In the NM approach the real part of the MOP is given by Eq.(\ref{eq5}), which in the case of Gogny force (\ref{eq9}) can easily be computed using the wave-functions (\ref{eq10}) and taking into account Eq.(\ref{eq11}). In the following we label the states of the projectile by the index $\alpha$ and the Fermi momentum of nucleons with the same (opposite) isospin as the projectile by  $\tau\alpha$  ($-\tau\alpha$), respectively. Thus, the real part of the MOP felt by an incident nucleon $\alpha$ in the NM approach computed with a Gogny force of the D1 family reads:

\setlength\mathindent{0pt}
\begin{eqnarray}
V_{\tau\alpha}=\frac{3}{2}t_3\rho^{1/3}[\rho-\rho_{\tau\alpha}]+\pi^{3/2}\sum_{k=1}^{k=2}\mu_k^3\Bigg[\left(W_k+\frac{B_k}{2}\right)\rho-\left(H_k+\frac{M_k}{2}\right)\rho_{\tau\alpha}\Bigg]-\nonumber\frac{1}{4\pi^{3/2}}\times\\ \sum_{k=1}^{k=2}\mu_k^3\Bigg[\left(\frac{W_k}{2}+B_k-\frac{H_k}{2}-M_k\right)I(k_\alpha,k_\rho=k_{\tau\alpha})-\left(\frac{H_k}{2}+M_k\right)I(k_\alpha,k_\rho=k_{-\tau\alpha})\Bigg] \label{eq12},
\end{eqnarray}
\setlength\mathindent{20pt}

where for each kind of particles its Fermi momentum is related to the corresponding particle density by $k_{\tau\alpha}^3=3\pi^2\rho_{\tau\alpha}$, while the momentum dependent functions $I$ introduced in Eq.(\ref{eq12}) are defined as

\begin{eqnarray}
I(k_\alpha,k_\rho)=\frac{4\pi^{3/2}}{\mu_k^3}\left[erf\left(\frac{\mu_k}{2}(k_\rho-k_\alpha)\right)+erf\left(\frac{\mu_k}{2}(k_\rho+k_\alpha)\right)\right]\nonumber \\ +\frac{8\pi}{k_\alpha\mu_k^4}\left[e^{-\frac{\mu_k^2}{4}(k_\alpha+k_\rho)^2}-e^{-\frac{\mu_k^2}{4}(k_\alpha-k_\rho)^2}\right] \label{eq13}.
\end{eqnarray}

Notice that Eq.(\ref{eq12}) corresponds to the strict HF potential, i.e. without rearrangement terms, computed in NM. To obtain the real part of the MOP in a finite nucleus we need to know how Eq.(\ref{eq12}) depends on the position $R$, which implies to know the radial dependence of the momentum of the projectile and of the density and Fermi momentum in the target nucleus. This is done  through the LDA as mentioned before. Therefore we assume that the real part of the optical potential in a finite nucleus at a given position $R$ is given by the same expression (\ref{eq12}) but with the densities and Fermi momenta replaced by the corresponding local values $\rho_{\tau\alpha}(R)$ and $k_{\tau\alpha}(R)$, respectively. These local values are obtained from a quasilocal HF calculation (see Appendix 2) performed in the target nucleus. The radial dependence of the momentum of the incident nucleon, $k_{\alpha}$ is given by the solution of the equation:

\begin{equation}
E_L = \frac{\hbar^2 k^2_{\alpha}}{2m} + V_{\tau\alpha}(k_{\alpha},k_{\tau\alpha}(R),k_{-\tau\alpha}(R))+ V_C (R)\delta_{\tau,1/2}, \label{eq14}
\end{equation}

where $E_L$ is the energy of the projectile in the laboratory frame and $V_C$ the Coulomb potential of the target, which contributes to the proton-nucleus processes. In Eq.(\ref{eq14}) the momentum of the incident nucleon $k_{\alpha}$ appears in both, kinetic and potential contributions, which implies that no analytical solution of (\ref{eq14}) exists in the case of Gogny forces and that the momentum $k_{\alpha}$ must be obtained numerically. However, the momentum of the incident nucleon can still be isolated by performing locally a quadratic Taylor expansion of the real part of the MOP $V_{\tau\alpha}$ around the Fermi momentum of the particles with the same isospin as the projectile $k_{\tau\alpha}$, i.e. 

\begin{equation}
V_{\tau\alpha}(R)=V_{\tau\alpha}(k_{\alpha}=k_{\tau\alpha},R)+ \left[\frac{1}{2k_{\alpha}}\frac{\partial V_{\tau\alpha}(k_{\alpha},R)}{\partial k_{\alpha}}\right]_{k_{\alpha}=k_{\tau\alpha}}(k_{\alpha}^2-k_{\tau\alpha}^2) \label{eq15},
\end{equation}

where in this equation the coordinate $R$ indicates the radial dependence of the real part of the MOP due to the local neutron and proton Fermi momenta.  Eq.(\ref{eq15}) can be recast as 

\begin{equation}
V_{\tau\alpha}(R)=V_{0\tau\alpha}(R)+b_{\tau\alpha}(R)k_{\alpha}^2 \label{eq16},
\end{equation}

with

\begin{equation}
V_{0\tau\alpha}=V_{\tau\alpha}(k_{\alpha}=k_{\tau\alpha})-\frac{k_{\tau\alpha}}{2}\left[\frac{\partial V_{\tau\alpha}}{\partial k_\alpha}\right]_{k_\alpha=k_{\tau\alpha}}, \label{eq17}
\end{equation}

and

\begin{equation}
b_{\tau\alpha}=\frac{1}{2k_{\tau\alpha}}\left[\frac{\partial V_{\tau\alpha}}{\partial k_{\alpha}}\right]_{k_\alpha=k_{\tau\alpha}} \label{eq18}.
\end{equation}

Eq.(\ref{eq18}) defines the inverse of the effective mass of the nucleons of the same type as the projectile computed at the Fermi momentum $k_{\tau\alpha}(R)$:

\begin{equation}
f_{\tau\alpha}(k_{\alpha}=k_{\tau\alpha},R) = \frac{m}{m^*_{\tau\alpha}} = 1 + \frac{2m}{\hbar^2}b_{\tau\alpha}(R) \label{eq19}.
\end{equation}

Thus in the potential defined by Eq.(\ref{eq16}) the momentum of the projectile appears explicitly isolated and the coefficients $V_{0\tau\alpha}$ and $b_{\tau\alpha}$, which  depend on the isospin $\tau$ of the projectile, are also functions of the position owing to their dependence on the Fermi momenta of neutrons and protons. Therefore the momentum of the projectile, as a function of the distance, can be obtained now from Eqs.(\ref{eq14}) and (\ref{eq16}) as 

\begin{equation}
k_{\alpha} = \sqrt{\frac{2m^*_{\tau\alpha}}{\hbar^2}\big(E_L - V_{0\tau\alpha}-  V_C \delta_{\tau,1/2}}\big) \label{eq20},
\end{equation}

which allows finally to write the radial dependence of the real part of the MOP in a finite nucleus calculated in the NM approach as

\begin{equation}
V_{\tau\alpha}= \frac{V_{0\tau\alpha}}{f_{\tau\alpha}} + (1 - \frac{1}{f_{\tau\alpha}})(E_L - V_C \delta_{\tau,1/2}) \label{eq21}.
\end{equation}

In Refs.\cite{pilipenko10,pilipenko12} the NM approach to the real part of the MOP was modified by including, on the one hand, the rearrangement term, which takes into account the modification of the mean-field due to the density-dependent term of the interaction, and, on the other hand, the contribution of the direct single-particle potential to the real part of the MOP is the one provided by the finite target nucleus instead of the one obtained in the homogeneous system. As was pointed out in Ref.\cite{pilipenko10}, to take the rearrangement term into account in the MOP may include, in an effective way, some diagrams of order equal or higher than three. In the case of Gogny forces the potential in finite nuclei, obtained introducing the aforementioned modifications in Eq.(\ref{eq12}), corresponds to the mean field of the target in the Slater approach and reads

\setlength\mathindent{0pt}
\begin{eqnarray}
&&V_{\tau\alpha}=\frac{3}{2}t_3\rho^{1/3}[\rho-\rho_{\tau\alpha}] + \frac{1}{4}t_3\rho^{-2/3}[\rho-\rho_{\tau\alpha}-\rho_{-\tau\alpha}]\nonumber \\ &&+\sum_{k=1}^{k=2}\left[\left(W_k+\frac{B_k}{2}\right)\int{d\vec{r}e^{-\frac{(\vec{R}-\vec{r})^2}{\mu^2_k}}\rho}-\left(H_k+\frac{M_k}{2}\right)\int{d\vec{r}e^{-\frac{(\vec{R}-\vec{r})^2}{\mu_k}}\rho_{\tau\alpha}}\right]-\frac{1}{4\pi^{3/2}}\nonumber \\ &&\sum_{k=1}^{k=2}\mu_k^3\left[\left(\frac{W_k}{2}+B_k-\frac{H_k}{2}-M_k\right) I(k_\alpha,k_\rho=k_{\tau\alpha})-\left(\frac{H_k}{2}+M_k\right)I(k_\alpha,k_\rho=k_{-\tau\alpha})\right] \label{eq22},
\end{eqnarray}
\setlength\mathindent{20pt}

which  we identify with the real part of the MOP. From now on, we refer to this approach to the real part of the MOP as HF approach. The choice of NM or HF approaches for describing the real part of the MOP determines the local behaviour of the momentum $k_{\alpha}$ of the projectile and has an important impact on the imaginary part of the MOP, as it can be seen in Figs. \ref{fig2} and \ref{fig3} that will be discussed in Section 3.2. The position dependence of the real part of the MOP in this HF approah is obtained in the same way as in using the NM prescription, as explained before. These two prescriptions, namely NM and HF, of the MOP based on the Gogny interaction, may be understood as two possible realizations of the optical potential within the inherent uncertainties of the optical model.  

\subsection{The imaginary part of the optical potential}

The imaginary part of the optical potential in the NM approach is obtained starting from Eq.(\ref{eq6}) and using the principal value integral formula, which is given by

\begin{equation}
\frac{1}{x+i\epsilon}=\frac{1}{x}-i\pi\delta(x). \label{eq23}
\end{equation}

To deal with the denominator of Eq.(\ref{eq6}), the parabolic approach to the potential (\ref{eq16}) is extremely useful because it allows to write the single particle energy of the projectile as 

\begin{equation}
\varepsilon_{\alpha}=\frac{\hbar^2}{2m}\big(1 +\frac{2m}{\hbar^2}b_{\tau\alpha}\big)k^2_{\alpha} + V_{0\tau\alpha}= \frac{\hbar^2f_{\tau\alpha}}{2m}k^2_{\alpha} + V_{0\tau\alpha}\label{eq24},
\end{equation}

and similar expressions for the remaining particle and hole states entering in Eq.(\ref{eq6}). Due to the energy conservation we can write the denominator of (\ref{eq6}) as 

\begin{equation}
\varepsilon_{\alpha} + \varepsilon_{\nu} -\varepsilon_{\lambda} - \varepsilon_{\mu} =\beta_{\tau\alpha}k^2_{\alpha} + \beta_{\tau\nu}k^2_{\nu} - \beta_{\tau\lambda}k^2_{\lambda}- \beta_{\tau\mu}k^2_{\mu} \label{eq25},
\end{equation}

where $\beta_{\tau\alpha}=\hbar^2f_{\tau\alpha}/2m$ and similarly for the other states that appear in Eq.(\ref{eq25}). After some lengthy algebra the imaginary part of the MOP in the NM approach calculated with the Gogny interaction can be expressed as:

\begin{equation}
W_{\tau\alpha}=-\frac{1}{2}\frac{\pi}{(2\pi)^6}\left[W_1+2W_2+W_3+W_4+W_5\right], \label{eq26}
\end{equation}

where the different contributions to (\ref{eq26}) are given by

\setlength\mathindent{0pt}
\begin{equation}
W_1=12t_3^2\rho^{2/3}I_1(\tau_\alpha,\tau_\nu=-\tau_\alpha), \label{eq27}
\end{equation}
\begin{eqnarray}
W_2=3t_3\rho^{1/3}\pi^{3/2}\sum_{k=1}^2\mu_k^3\bigg[W_k+B_k+H_k+M_k\bigg]\bigg[I_2(\tau_\alpha,-\tau_\alpha) +I_3(\tau_\alpha,-\tau_\alpha)\bigg], \label{eq28}
\end{eqnarray}
\begin{eqnarray}
W_3=\pi^3\sum_{k=1}^2\mu_k^6\Big\{[W_k(2W_k+B_k)+B_k(2B_k+W_k)+H_k(2H_k+M_k)+M_k(2M_k+H_k)]&&\nonumber \\  \times [I_4(\tau_\alpha,\tau_\alpha)+I_5(\tau_\alpha,\tau_\alpha)+I_4(\tau_\alpha,-\tau_\alpha)+I_5(\tau_\alpha,-\tau_\alpha)]
&&\nonumber \\-2[W_k(2H_k+M_k)+B_k(2M_k+H_k)][I_4(\tau_\alpha,\tau_\alpha)+I_5(\tau_\alpha,\tau_\alpha)] &&\nonumber  \\ -2[W_k(2B_k+W_k)+B_k(2W_k+B_k)+H_k(2M_k+H_k)+M_k(2H_k+M_k)]I_6(\tau_\alpha,\tau_\alpha)  &&\nonumber \\ +4[W_k(2M_k+H_k)+B_k(2H_k+M_k)][I_6(\tau_\alpha,\tau_\alpha)+I_6(\tau_\alpha,-\tau_\alpha)]\Big\}, && \\ \nonumber \label{eq29}
\end{eqnarray}
\begin{eqnarray}
&&W_4=\pi^3 \sum_{i,j=1 \atop i\neq j}^2\mu_I^3\mu_J^3\Big\{[I_7(\tau_\alpha,\tau_\alpha)+I_8(\tau_\alpha,\tau_\alpha)+ I_7(\tau_\alpha,-\tau_\alpha)+I_8(\tau_\alpha,-\tau_\alpha)] [W_i(2W_j+B_j)+\nonumber \\ &&B_i(2B_j+W_j)+H_i(2H_j+M_j)+M_i(2M_j+H_j)]-\nonumber \\&&[I_7(\tau_\alpha,\tau_\alpha) +I_8(\tau_\alpha,\tau_\alpha)][W_i(2H_j+M_j) +B_i(2M_j+H_j)+W_j(2H_i+M_i)+\nonumber \\ &&B_j(2M_i+H_i)]-[I_9(\tau_\alpha,\tau_\alpha)+I_{10}(\tau_\alpha,\tau_\alpha)]
[W_i(2B_j+W_j)+B_i(2W_j+B_j)+\nonumber \\ &&H_i(2M_j+H_j)+ M_i(2H_j+M_j)]+[I_9(\tau_\alpha,\tau_\alpha)+I_{10}(\tau_\alpha,\tau_\alpha)+I_9(\tau_\alpha,-\tau_\alpha)+\nonumber \\ &&I_{10}(\tau_\alpha,-\tau_\alpha)][W_i(2M_j+H_j)+B_i(2H_j+M_j)+W_j(2M_i+H_i)+B_j(2H_i+M_i)]\Big\} \label{eq30},
\end{eqnarray}

and 

\begin{equation}
W_5=\frac{1}{2}W_{LS}^2\left[2I_{11}(\tau_\alpha,\tau_\alpha)+I_{11}(\tau_\alpha,-\tau_\alpha)\right] \label{eq31}.
\end{equation}

All these contributions to the imaginary part of the MOP depend on the parameters of the interaction (see Eq.(\ref{eq9})) as well as on the integrals $I_i(\tau_\alpha,\tau_\nu)(\tau_\nu=\tau_\alpha,-\tau_\alpha)$ with $i=1 \dots 10$, which are defined as

\begin{equation}
I_i(\tau_\alpha,\tau_\nu)=\int d\vec{k}_\nu d\vec{k}_\lambda d\vec{k}_\mu f_i(\gamma_1,\gamma_2)\delta(\vec{k}_\alpha+\vec{k}_\nu-\vec{k}_\lambda-\vec{k}_\mu)\delta(\beta_{\tau\alpha}k_\alpha^2+\beta_{\tau\nu}k_\nu^2-\beta_{\tau\alpha}k_\lambda^2-\beta_{\tau\nu}k_\mu^2). \label{eq32}
\end{equation}
\setlength\mathindent{20pt}

They are explicitly calculated in the Appendix 1. \\

The functions $f_i(\gamma_1,\gamma_2)$, which appear in the integrands of $I_i(\tau_\alpha,\tau_\nu)$, are defined as $f_i(\gamma_1,\gamma_2)=e^{-\frac{\gamma_1^2}{4}k_{\alpha\lambda}^2-\frac{\gamma_2^2}{4}k_{\alpha\mu}^2}$ with $\vec{k}_{\alpha\lambda}=\vec{k}_\alpha-\vec{k}_\lambda$ and $\vec{k}_{\alpha\mu}=\vec{k}_\alpha-\vec{k}_\mu$. These functions for the indices $i=1\dots 10$ correspond to different values of the arguments $\gamma_1,\gamma_2$ and are given by \\

\begin{tabular}{llllll}
$f_1$&=&$f(\gamma_1=0,\gamma_2=0)$, & $f_2$&=&$f(\gamma_1=\mu_k,\gamma_2=0)$, \\ $f_3$&=&$f(\gamma_1=0,\gamma_2=\mu_k)$, &$f_4$&=&$f(\gamma_1=\sqrt{2}\mu_k,\gamma_2=0)$, \\ $f_5$&=&$f(\gamma_1=0,\gamma_2=\sqrt{2}\mu_k)$, &$f_6$&=&$f(\gamma_1=\mu_k,\gamma_2=\mu_k)$, \\ $f_7$&=&$f(\gamma_1=\sqrt{\mu_i^2+\mu_j^2},\gamma_2=0)$, & $f_8$&=&$f(\gamma_1=0,\gamma_2=\sqrt{\mu_i^2+\mu_i^2})$, \\ $f_9$&=&$f(\gamma_1=\mu_i,\gamma_2=\mu_j)$, &$f_{10}$&=&$f(\gamma_1=\mu_j,\gamma_2=\mu_i)$. \\ 
\end{tabular} \\

Finally, the function $f_{11}$ entering in the spin-orbit contribution to the central imaginary part of the MOP in the NM approach (\ref{eq31}) is given by $f_{11} = \big(\vec{k}_{\alpha\nu} \times \vec{k}_{\lambda\mu}\big)^2$ \cite{shen09,xu14} where, as before, $\vec{k}_{\alpha\nu}=\vec{k}_\alpha-\vec{k}_\nu$ and $\vec{k}_{\lambda\mu}=\vec{k}_\lambda-\vec{k}_\mu$. \\

In the NM approach the imaginary part of the MOP is given in terms of the integrals $I_i(\tau_\alpha,\tau_\nu)$, which are fully determined by the momentum of the projectile $k_{\alpha}$ and of the Fermi momenta of neutrons and protons in the target. In a finite nucleus all these momenta become functions of the distance through the LDA, as we have explained in the previous subsection 2.1, which allows to write the imaginary part of the MOP as a function of the radial coordinate $R$.

\section{Results and discussions}

In this section we  analyze different scattering observables computed with the Gogny MOP introduced in the previous section. It is important to note that our model, in its present version, has no free parameters fitted to reproduce scattering data and, therefore, is fully predictive. First we discuss the effective mass approach, which is essential to obtain the central imaginary part of the optical potential in our model in a relatively simple way. Next the central real and imaginary parts as well as the real spin-orbit contribution predicted by the Gogny MOP are compared with the corresponding terms of the global POP fitted by Koning and Delaroche (POPKD from now on). In the third subsection we compare the differential cross sections (DCS) and analyzing powers (AP) obtained with our model with the experimental data and with the predictions of the POPKD for a sample of elastic neutron and proton reactions on different targets at several energies. We also analyze as a function of the energy of the projectile the total (for neutrons) and reaction (for protons) cross sections as well as the volume integrals of the real and imaginary parts of the central contribution to the MOP for neutrons. Finally, in order to check the predictive power of our model we compare our results for neutron and proton elastic scattering on $^{40}$Ca and $^{48}$Ca with the ones computed by  Blanchon and collaborators with the NSM \cite{blanchon15a,blanchon15b,blanchon17}.

\subsection{The parabolic approach to the momentum dependence of the real part of the optical potential} 

\begin{figure}[ht]
\centering
\includegraphics[width=1.0\linewidth,clip=true]{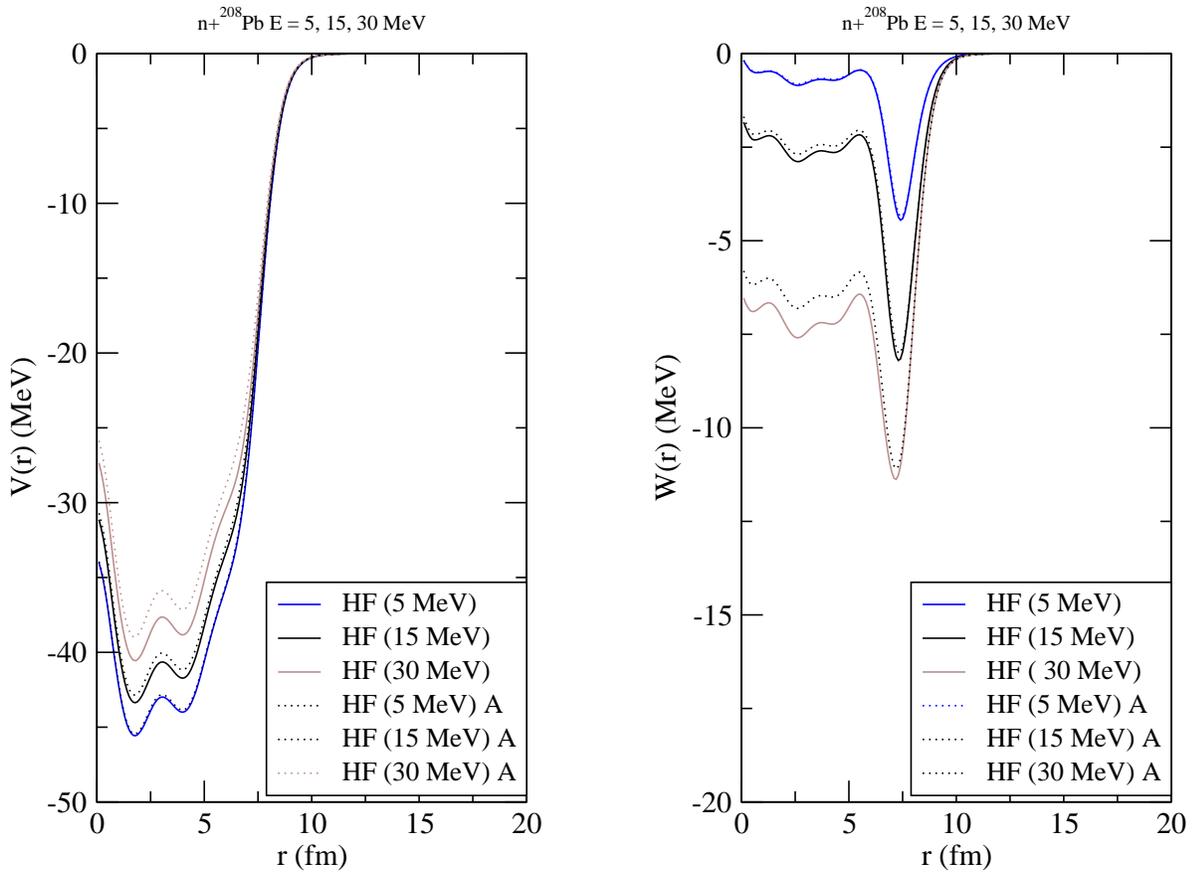}
\setlength\mathindent{20pt}
\caption{Real and imaginary parts of the central contribution to the MOP of $^{208}$Pb calculated with the HF prescription for neutrons with energies of 5, 15 and 30 MeV. For each energy the solid lines display the result obtained with the exact real part (\ref{eq22}) and the dashed line corresponds to its parabolic approach (\ref{eq15}).}
\label{fig1}
\end{figure}

As mentioned before, an essential approximation, which is needed for computing the imaginary part of the MOP in an almost analytical way, is to simplify the involved dependence of the real part of the MOP on the momentum of the  projectile. To this end we perform a quadratic Taylor expansion of $k_{\alpha}$ around its Fermi momentum in Eq.(\ref{eq22}). To check the accuracy of this expansion, we plot in the left panel of Figure \ref{fig1} the exact real part of the MOP estimated with the HF prescription (\ref{eq22}) (solid line), and compare with the corresponding parabolic approach given by (\ref{eq15}) (dashed line), for neutrons of 5, 15 and 30 MeV scattered by a target nucleus $^{208}$Pb. From this panel we can see that the exact and the parabolic approximated MOP in the HF estimate agree at the surface but differ slightly in the bulk region, with differences that are  a little bit larger when the energy of the projectile increases. In the right panel of the same Figure \ref{fig1} we also compare the imaginary part of the MOP (\ref{eq26}) calculated with the radial dependence of the momentum of the projectile $k_{\alpha}({\bf R})$ obtained  from the real part of the MOP as the solution of Eq.(\ref{eq14}) at HF level (\ref{eq22}) (solid lines) and computed using the corresponding parabolic approach (\ref{eq15}) (dashed lines). From this panel we see that, similarly to the real part, the agreement between the exact and parabolic approximated imaginary parts of the MOP is again very good at the surface and differs a little more in the bulk, where the differences are larger when the energy of the incident neutron grows. 

We have also checked that if the NM approach (\ref{eq12}) is used to obtain the radial dependence of momentum of the projectile instead of the HF one, the real and imaginary parts of the MOP, computed exactly or through the parabolic approach, show qualitatively a similar behavior to that exhibited in the HF case as displayed in Figure \ref{fig1}. These results show that the parabolic approach for dealing with the momentum dependence of the real part of the MOP is very reliable and can be applied confidently in order to compute the imaginary part of the MOP. \\ \\ 

\setlength\mathindent{20pt}
\begin{figure}[ht]
\centering
\includegraphics[width=1.0\linewidth,clip=true]{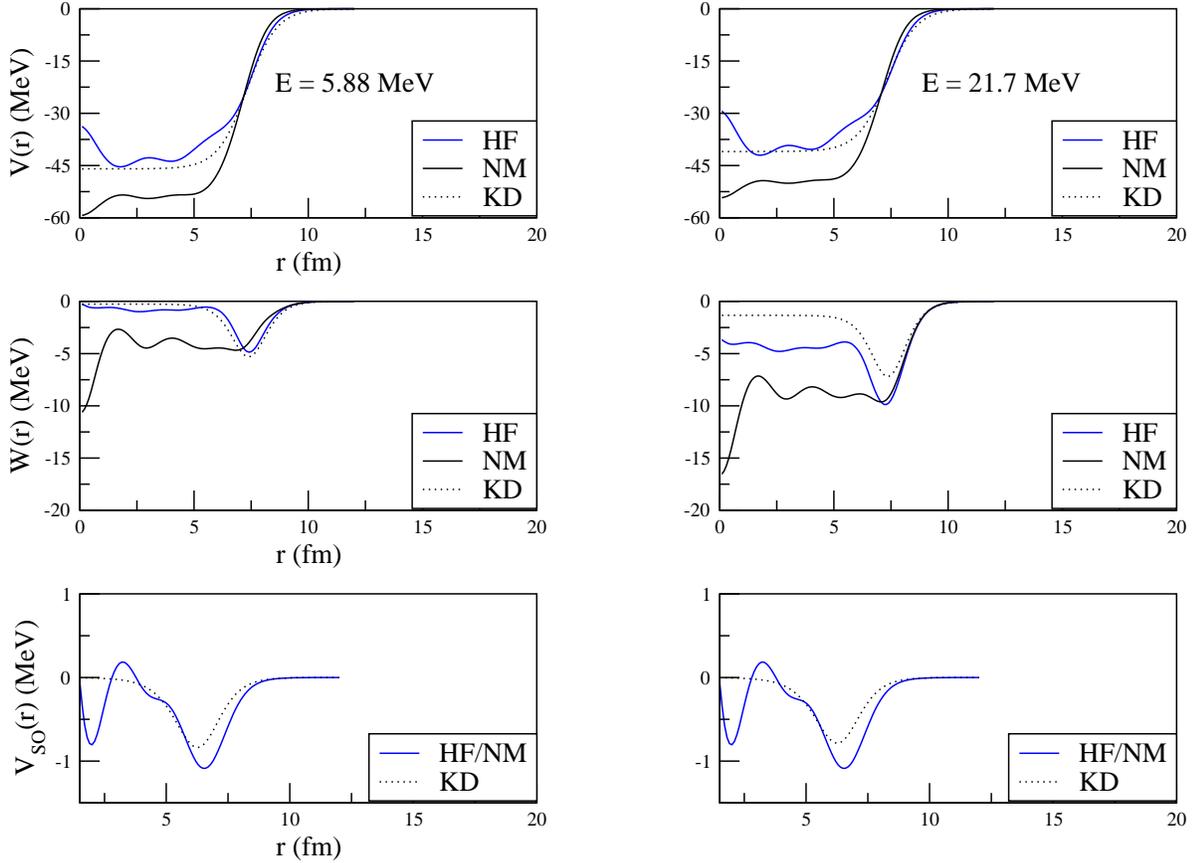}
\setlength\mathindent{20pt}
\caption{Real (upper panel) and imaginary (middle panel) contributions to the central part and the real spin-orbit term (lower panel) of the MOP for neutrons of 5.88/21.7 MeV scattered by $^{208}$Pb.}
\label{fig2}
\end{figure}
\setlength\mathindent{20pt}

\begin{figure}[hb]
\centering
\includegraphics[width=1.0\linewidth,clip=true]{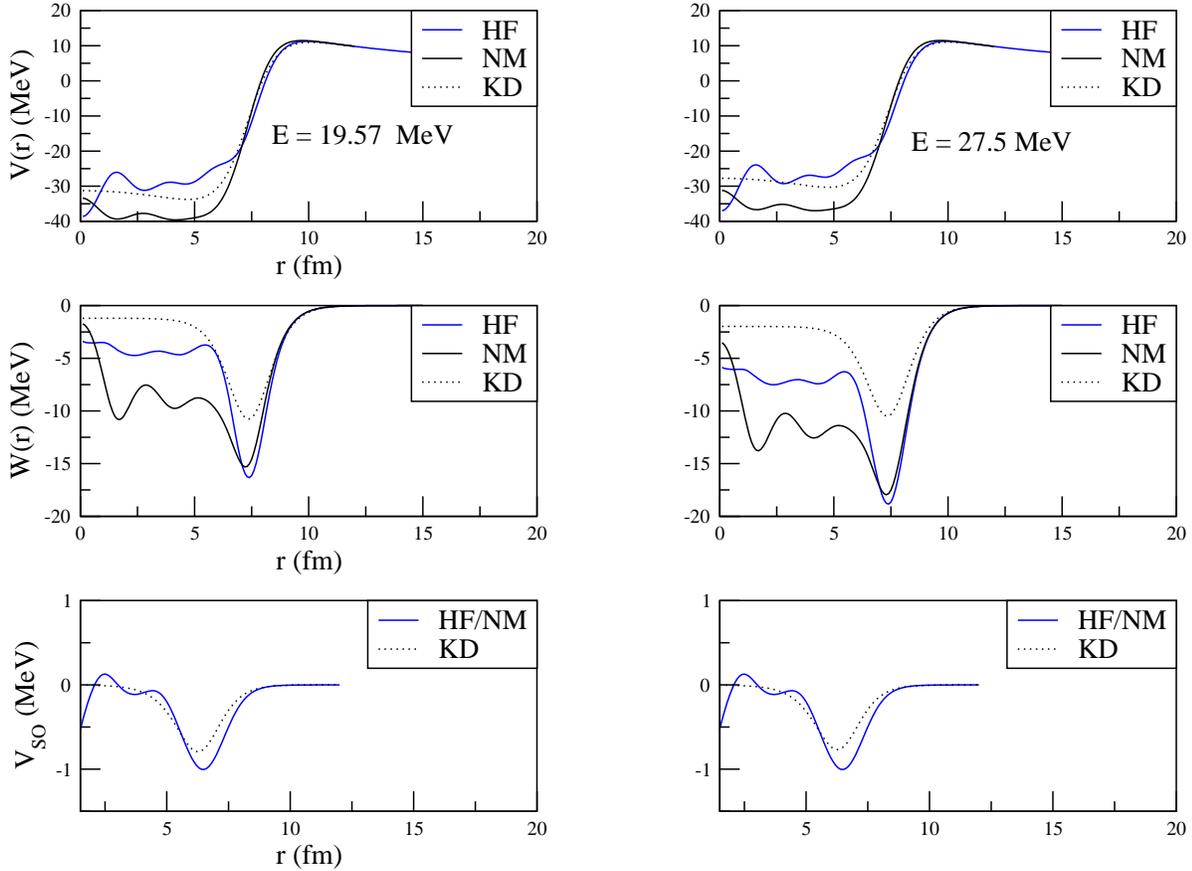}
\setlength\mathindent{20pt}
\caption{Real (upper panel) and imaginary (middle panel) contributions to the central part and the real spin-orbit term (lower panel) of the MOP for protons of 19.57/27.5 MeV scattered by $^{208}$Pb}
\label{fig3}
\end{figure}
\setlength\mathindent{20pt}

\subsection{The MOP potential in the nuclear matter approach calculated with Gogny forces}

In the first and second panels of each column of Figures \ref{fig2} and \ref{fig3} we display the NM and HF estimates of the real and imaginary parts of the central contribution to the MOP calculated using the Gogny D1S force for elastic scattering of neutrons (Figure \ref{fig2}) and protons (Figure \ref{fig3}) on $^{208}$Pb at several energies of the projectiles. For comparison we also display in these Figures the same quantities computed with the global POPKD \cite{koning03}. For the real part of the central term of the MOP we see that the POPKD lies in between the HF and NM estimates of our Gogny MOP, for both neutron and proton projectiles. As a general trend, the real part obtained using the HF estimate is closer to the POPKD compared to that  using the NM estimate  in the case of neutron scattering while the contrary happens when the projectiles are protons. The imaginary part of the central term of the MOP shows similar behavior for both neutron and  proton scattering. We see that for both types of projectile the imaginary part obtained with the HF estimate has a surface contribution more important than the volume part, which is in  qualitative agreement with the behavior shown by the POPKD. However, if the NM estimate is used, the imaginary central part shows an important volume contribution as compared with the surface one. Although the position of the surface peaks in the imaginary part of the central term obtained with our model for both types of projectiles using both the NM and HF estimates coincides well with the predictions of the POPKD, the depth of the imaginary term differ more between both calculations. We see that our calculation gives more depth to the single-particle potential and the difference with the POPKD increases when the energy of the projectile grows, pointing out a different energy dependence of our Gogny MOP compared with the POPKD behavior. In the bottom-most panels of the same Figures we display our estimate of real part of the spin-orbit contribution to the MOP. It is obtained from the self-consistent quasi-local HF calculation \cite{gridnev98,soubbotin00,soubbotin03} in the target nucleus as is discussed in Appendix 2. From these panels we see that the real part of the spin-orbit term calculated using the Gogny D1S force is in good agreement with the corresponding part of the POPKD.  We have also checked that the shapes displayed by the real and imaginary parts of the central term of the MOP as well as its spin-orbit contribution for scattering of neutron and proton by nuclei analyzed here in the particular case of $^{208}$Pb are similar if  different other targets are considered, with the main difference that the position of the surface varies as a function of $A^{1/3}$.

\subsection{Neutron and proton elastic scattering on $^{56}$Fe}

\begin{figure}[ht]
\centering
\includegraphics[width=1.0\linewidth,clip=true]{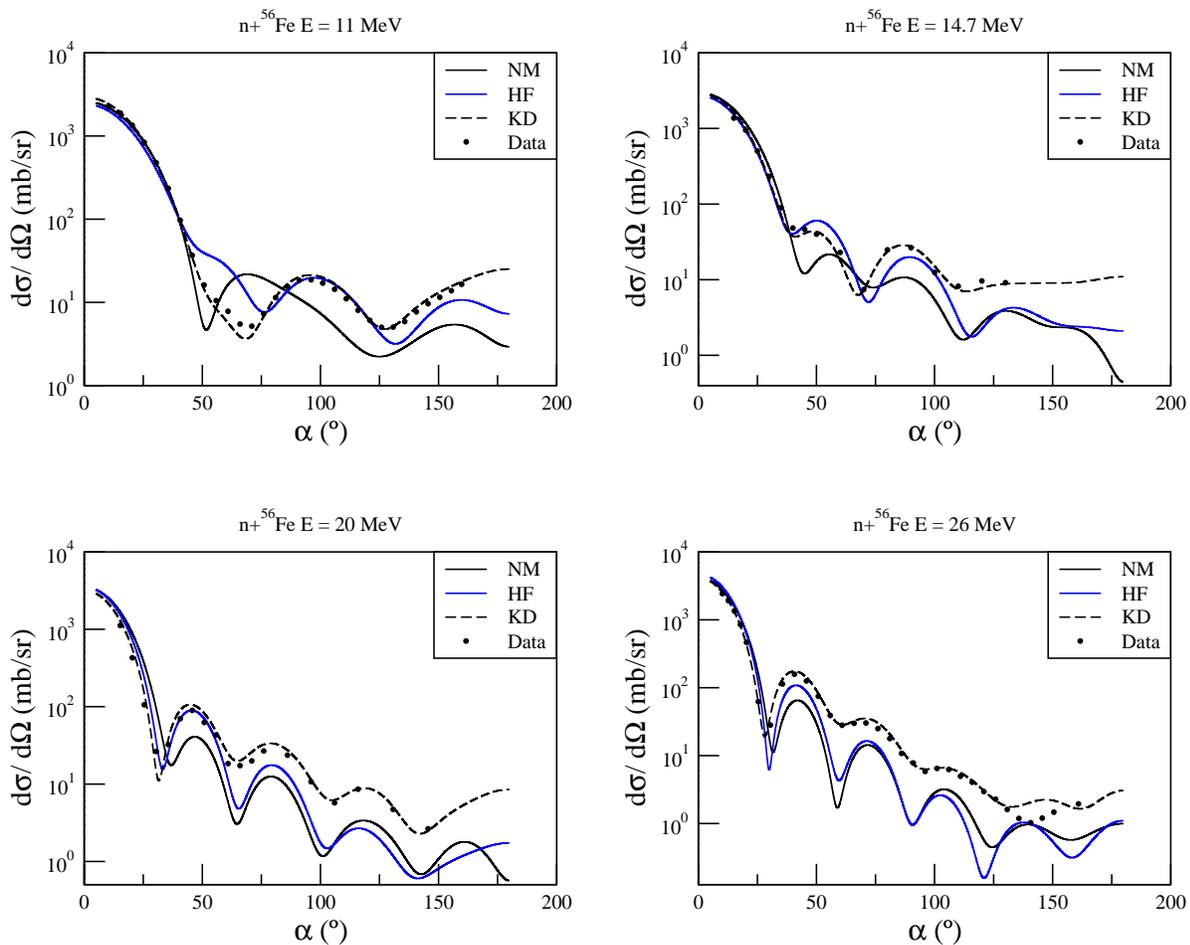}
\setlength\mathindent{20pt}
\caption{Differential cross section of scattering of neutrons at several energies off  $^{56}$Fe}
\label{fig4}
\end{figure}

\begin{figure}[hb]
\centering
\includegraphics[width=1.0\linewidth,clip=true]{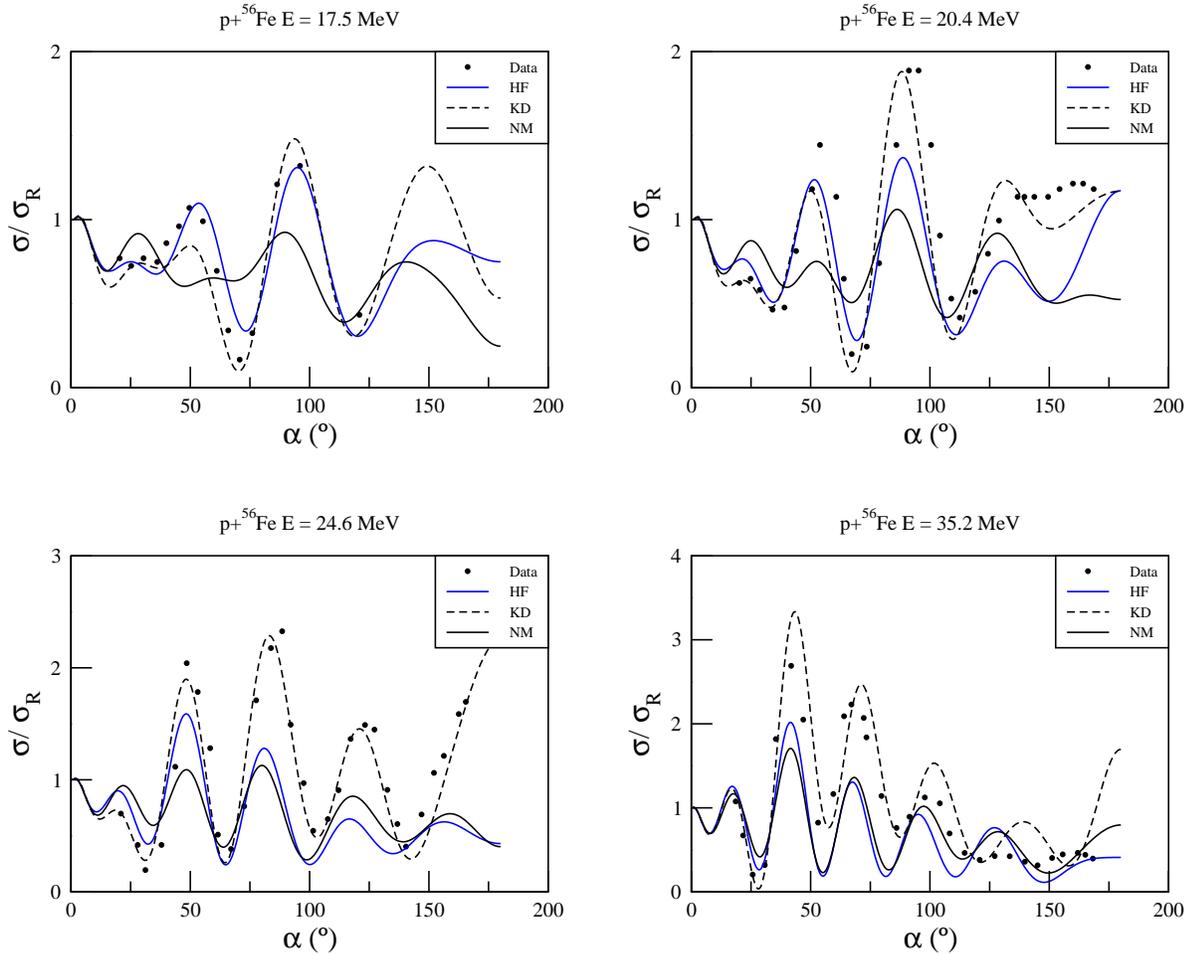}
\setlength\mathindent{20pt}
\caption{Differential cross section of scattering of protons at several energies off $^{56}$Fe}
\label{fig5}
\end{figure}

In this subsection we want to investigate the predictive power of the Gogny MOP discussed in the previous sections and computed using the D1S Gogny force to describe scattering observables. To this end, we display in Figures \ref{fig4} and \ref{fig5} the DCS for neutron and proton elastic scattering by $^{56}$Fe at energies between 11 and 26 MeV for neutrons and between 17.5 and 35.2 MeV for protons. We show the results obtained with our model computing the real and imaginary parts of the central term with the NM (Eq.\ref{eq12}) and the HF (Eq.\ref{eq22}) prescriptions. In the same figure, and as a benchmark, we also show the DCS for the same reactions calculated with the POPKD as well as the experimental data taken from the EXFOR web-page \cite{exfor}. In spite of the fact that $^{56}$Fe is slightly deformed \cite{cea}, it has been shown that the elastic DCS data for neutron and proton up to energies of 65 MeV can be fairly well described using the modified JLM MOP of Ref.\cite{bauge98}. We want to study the predictions of our Gogny MOP for this nucleus, because here exists a lot of scattering data, which allows us to study for a given target the projectile energy dependence in our model. In our analysis we disregard elastic neutron reactions with bombarding energies below $\sim$ 10 MeV because our model does not take into account the compound nucleus contribution (see \cite{blanchon15b} for more details). We can see that in the case of neutron reactions and for small scattering angles up to about 50$^{o}$ the DCS evaluated with the Gogny MOP, in particular for the HF approach, are in reasonably good agreement with the prediction of the POPKD as well as with the experimental data. For larger scattering angles the Gogny MOP shows a  pattern similar to that of the POPKD, but with stronger dips, which are roughly located at the same scattering angles predicted by the POPKD. \\

In Figure \ref{fig5} we display the ratio of the proton Gogny MOP DCS to Rutherford DCS. For this nucleus and for the range of the incident energies considered in this work, the experimental data show an oscillatory pattern as a function of the scattering angle with amplitudes that are damped for backwards angles when the incident energy is high enough. The Gogny MOP values, in particular the ones obtained using the HF prescription, follow reasonably well the experimental trends but with relatively small amplitudes, as compared with the experimental data and the POPKD results. The largest differences between the Gogny MOP predictions and the experimental data and POPKD values are concentrated in backwards angles larger than 150$^{o}$ as can be seen in Figure \ref{fig5}.

\subsection{Neutron and proton elastic scattering in medium size targets.}

In Fig.\ref{fig6} we display the DCS of the reactions n+$^{98}$Mo at 11 and 26 MeV and n+$^{90}$Zr at 11 and 24 MeV. The DCS computed with the Gogny MOP of this work show again a similar pattern to that predicted by the POPKD. The agreement between our results and the POPKD DCS as well as the experimental data is fairly good for forward angles up to around 40$^{o}$ and 25$^{o}$ for low and high bombarding energies, respectively. The position of the  dips of the DCS computed with our Gogny MOP are shifted with respect to the predictions of the POPKD and the experimental values towards larger (with the HF approach) or smaller (with the NM approach) scattering angles, respectively. Notice, however, that this shift is reduced when the energies of the bombarding neutrons increase, although the absorption predicted by our model is, in general, comparably too strong, in particular at high energies, as is seen from the lowest panels of Figure \ref{fig6}. \\

Figure \ref{fig7} displays the ratio of the proton DCS to Rutherford DCS for the reactions p+$^{58}$Ni at 20.4 and 40 MeV and p+$^{90}$Zr at 25.05 and 49.35 MeV. From this Figure we see that  the predictions of the Gogny MOP reproduce the oscillatory behavior exhibited by the experimental data, although the amplitude of the oscillations is damped as compared with the predictions of the POPKD and the experimental data. At relatively low energy the HF prescription for the Gogny MOP agrees better with the POPKD results than those from the NM approach. At higher incident energy both HF and NM prescriptions of the Gogny MOP predict almost the same values reproducing fairly well the experimental data and  follow better the POPKD trends than at low energy.

\setlength\mathindent{20pt}
\begin{figure}[hb]
\centering
\includegraphics[width=0.8\linewidth,clip=true]{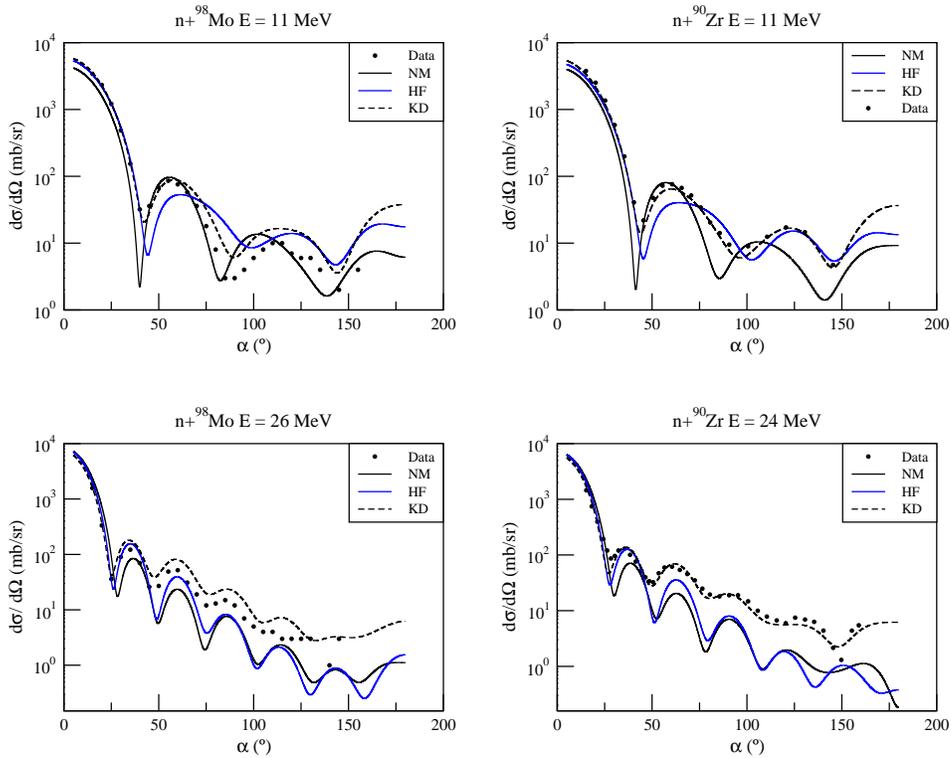}
\setlength\mathindent{20pt}
\caption{Differential cross section of scattering of neutrons at several energies off $^{98}$Mo, $^{90}$Zr}
\label{fig6}
\end{figure}
\setlength\mathindent{20pt}

\begin{figure}[hb]
\centering
\includegraphics[width=1.0\linewidth,clip=true]{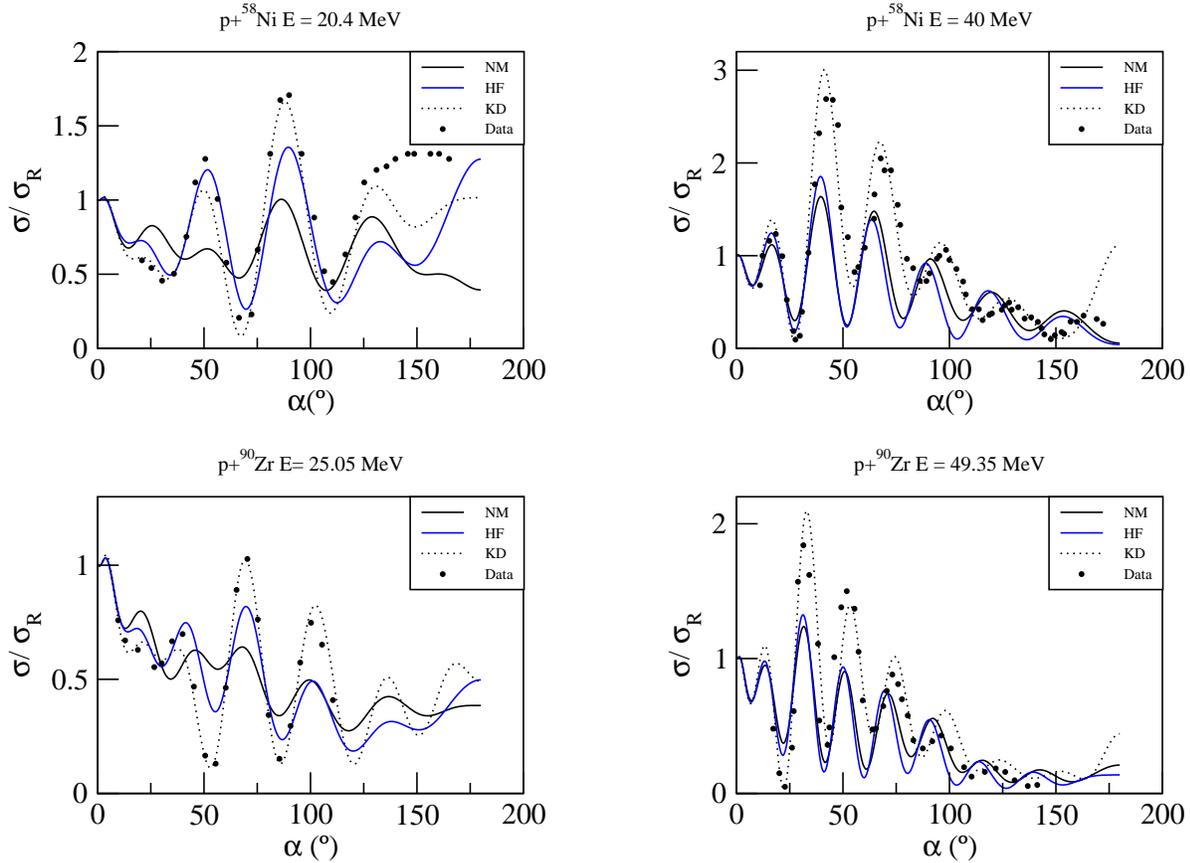}
\setlength\mathindent{20pt}
\caption{Differential cross section of scattering of protons at several energies off $^{66}$Zn and $^{90}$Zr}
\label{fig7}
\end{figure}

\subsection{Neutron and proton elastic differential cross sections in heavy targets.}

In Figure \ref{fig8} we display the DCS for neutron (upper panels) and proton (lower panels) elastic scattering on a target of $^{208}$Pb at different bombarding energies. For neutron scattering the DCS behave in a rather similar way as in the case of lighter targets. The Gogny MOP results agree reasonably well with the experimental data and with the POPKD predictions for forward angles up to about 50$^o$. For larger angles the differences are more relevant. On the one hand, the positions of the dips in the DCS predicted by the Gogny MOP, with both NM and  HF prescriptions, are slightly shifted with respect to the experimental and POPKD dips. On the other hand, the strength of the DCS for large scattering angles predicted by the Gogny MOP is smaller than the experimental and POPKD values, which implies that our model provides an imaginary central potential too strong, specially when the energy of the neutrons increases (see also in this respect Figure \ref{fig2}). \\

For proton reactions on $^{208}$Pb the situation is different. For the two analyzed energies, the experimental values of the ratio of the proton DCS to Rutherford DCS show an oscillatory decreasing trend with the scattering angle, which is well reproduced by the POPKD but not by the Gogny MOP. In this case although $\sigma/\sigma_R$ decreases with the angle, its oscillations are almost completely suppressed and its value is slightly larger than the experimental data and the POPKD predictions.    

\setlength\mathindent{20pt}
\begin{figure}[hb]
\centering
\includegraphics[width=1.0\linewidth,clip=true]{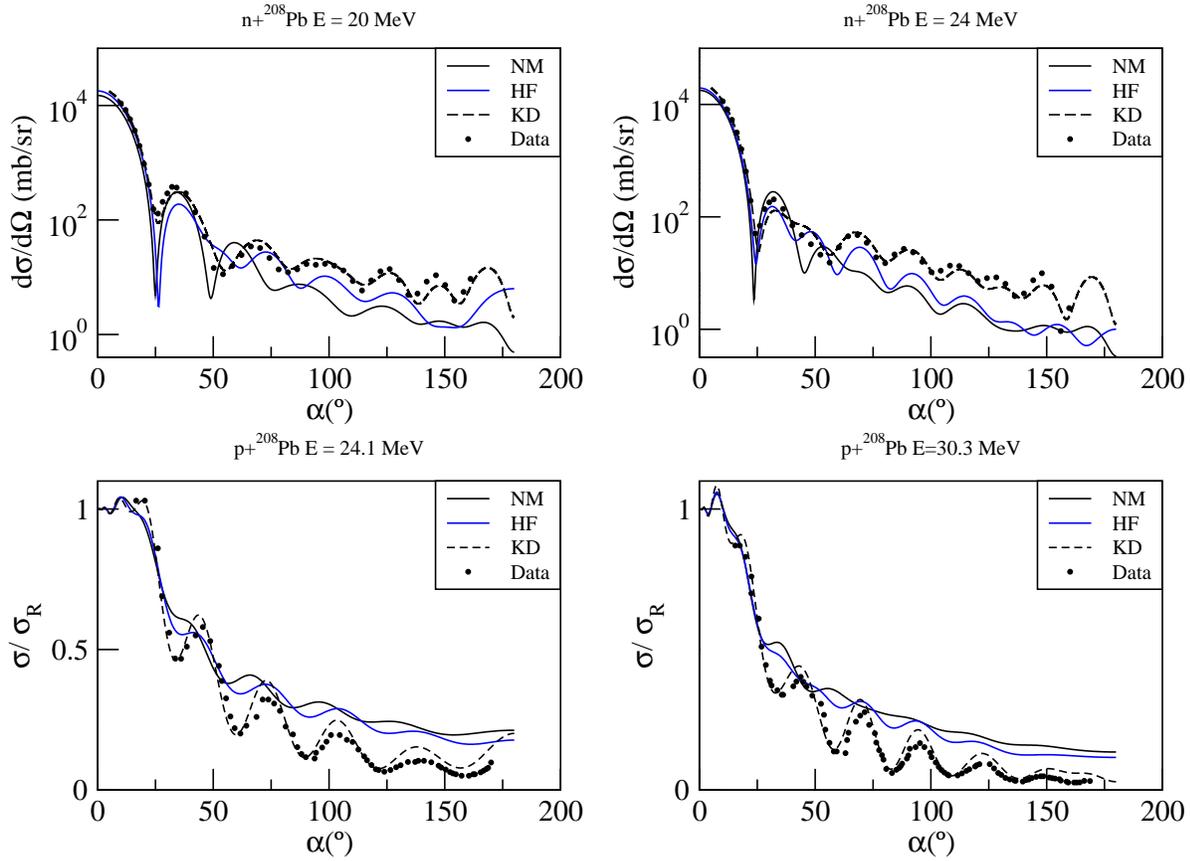}
\setlength\mathindent{20pt}
\caption{Differential cross section of scattering of neutrons and protons at several energies off $^{208}$Pb.}
\label{fig8}
\end{figure}
\setlength\mathindent{20pt}

\subsection{Analyzing power}
\setlength\mathindent{20pt}
\begin{figure}[hb]
\centering
\includegraphics[width=1.0\linewidth,clip=true]{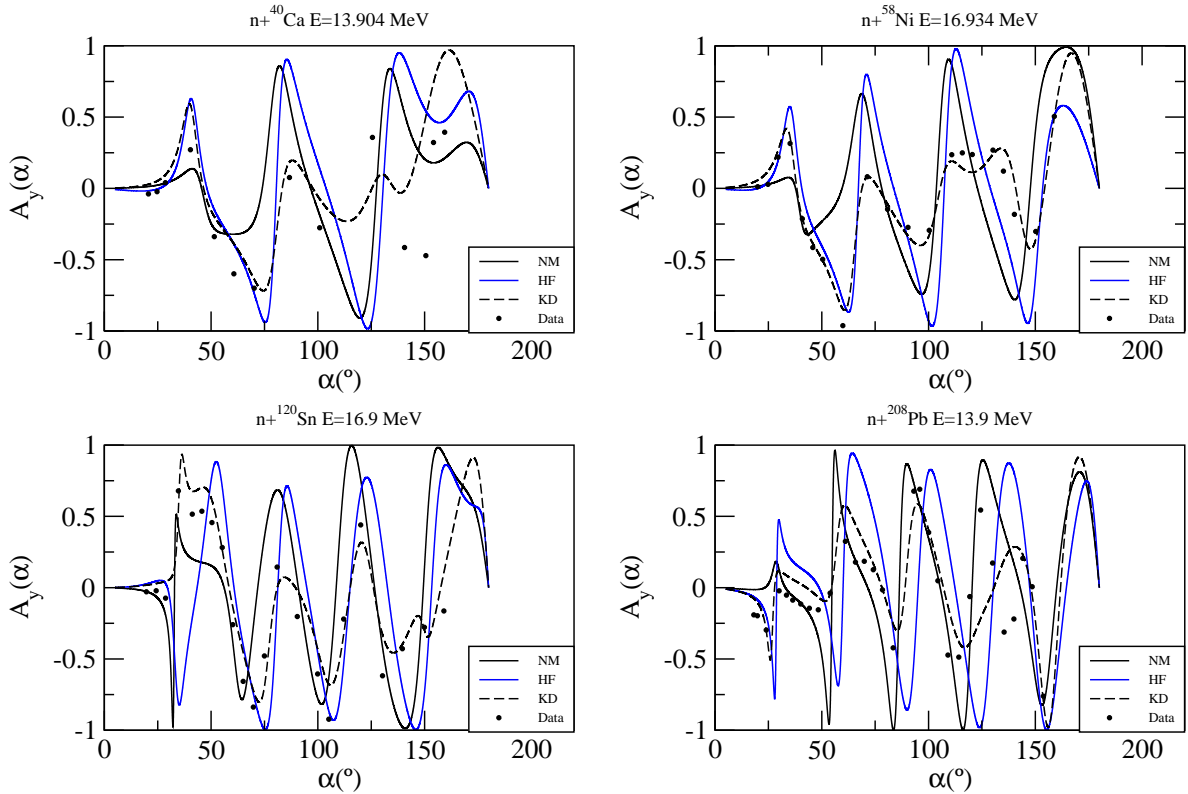}
\setlength\mathindent{20pt}
\caption{Analyzing powers for neutron scattering off $^{40}$Ca, $^{58}$Ni, $^{120}$Sn and $^{208}$Pb at several energies}
\label{fig9}
\end{figure}

\begin{figure}[hb]
\centering
\includegraphics[width=1.0\linewidth,clip=true]{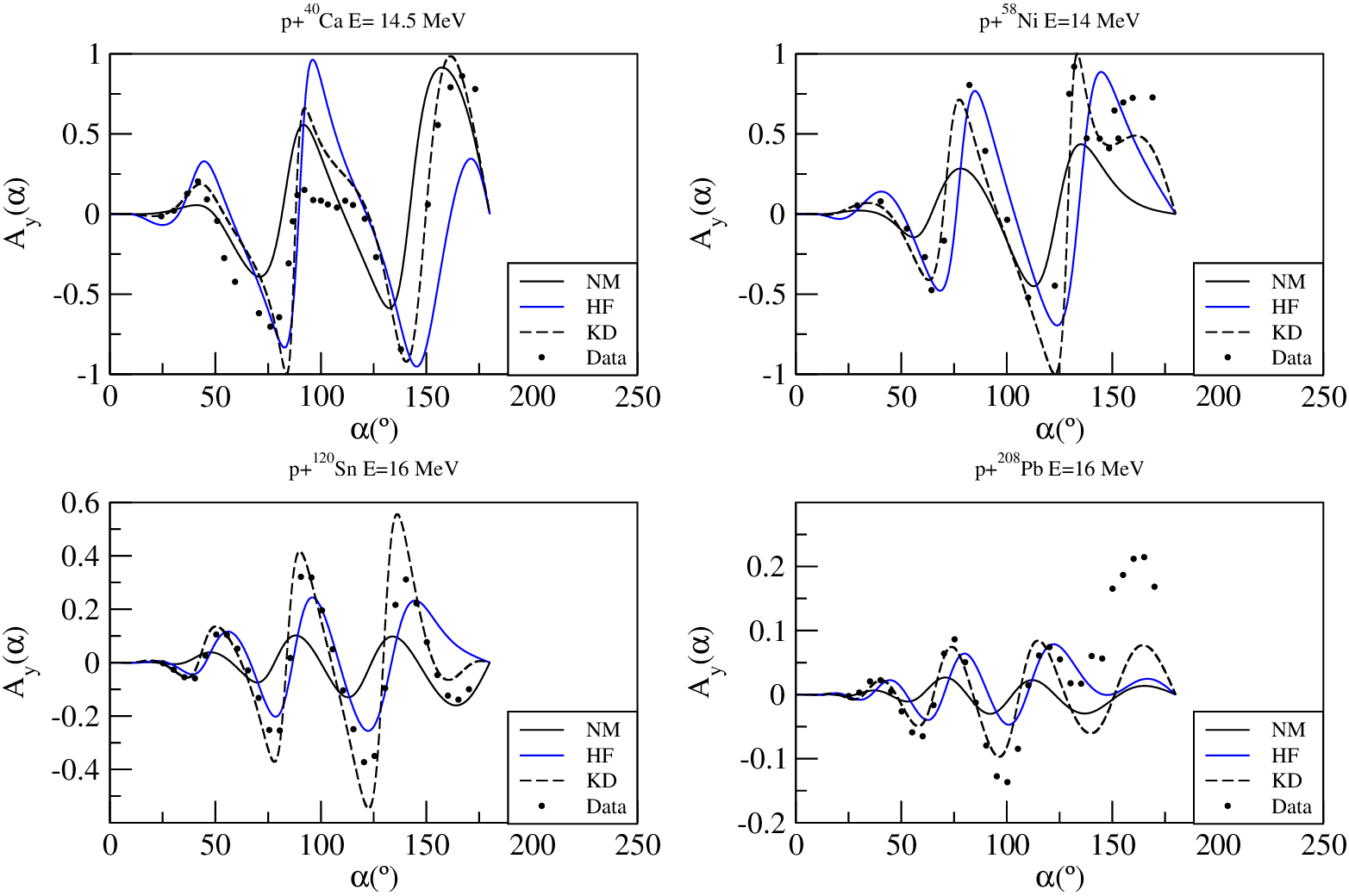}
\setlength\mathindent{20pt}9
\caption{Analyzing powers for proton scattering off $^{40}$Ca, $^{58}$Ni, $^{120}$Sn and $^{208}$Pb at several energies}
\label{fig10}
\end{figure}

In Figures \ref{fig9} and \ref{fig10} we display the analyzing power (AP) of neutron (upper panels) and proton (lower panels) elastic scattering at several energies on targets of $^{40}$Ca, $^{58}$Ni, $^{120}$Sn and $^{208}$Pb. In the same Figures we also display the experimental data as well as results from the POPKD predictions. For neutrons we can see that the Gogny MOP AP, obtained with both NM and HF prescriptions, follow rather well the oscillations predicted by the POPKD and reproduce the experimental data, in particular for light nuclei and for scattering angles less than 50$^o$. In the case of protons and again for light nuclei, the AP computed with the Gogny MOP reproduces the experimental data as well as the POPKD results quite well, even for backward scattering angles, specially using the HF prescription. However, the Gogny MOP description of the AP worsens in the case of medium and heavy targets. Using the NM prescription the oscillations of the AP are roughly in phase with the ones of POPKD although its amplitudes are clearly damped in the Gogny MOP case as compared with the POPKD values. Using the HF prescription, the oscillations in the AP are shifted  with respect to the position predicted by the POPKD and the amplitudes of the oscillations grow for backwards angles.

\setlength\mathindent{20pt}

\subsection{Global properties}

In the previous subsections we have compared some local observables like  the differential cross sections and analyzing powers obtained with the Gogny MOP developed in this work with the predictions of the POPKD (which we use as a benchmark), as well as with the experimental data. In this section we want to discuss other results that show the quality of our predictions  and allow to get a more global insight about the description of nucleon-nucleus reactions provided by our model.\\

\begin{table}
\begin{center}
\begin{tabular}{|c|c|c|c|c|c|c|c|}
\hline
\multicolumn{4}{|c|}{$^{56}$Fe(N,EL)$^{56}$Fe} & \multicolumn{4}{c|}{$^{56}$Fe(P,EL)$^{56}$Fe}\\
\hline
E(MeV) & KD & HF & NM & E(MeV) & KD & HF & NM \\
\hline
17.5 & 0.2209 & 0.3881 & 0.8566 &  11 & 0.1265 & 0.7850 & 0.9457 \\
\hline
20.4 & 0.3216 & 0.3439 & 0.5609 & 14.7 & 0.1337 & 0.3756 & 0.5046 \\
\hline
24.6 & 0.2687 & 0.5121 & 0.6190 & 20 & 0.2388 & 0.5691 & 1.0076 \\
\hline
35.2 & 0.5011 & 0.4966 & 0.5126 & 26 & 0.2866 & 0.6008 & 0.6446 \\
\hline
\multicolumn{4}{|c|}{$^{98}$Mo(N,EL)$^{98}$Mo} & \multicolumn{4}{c|}{$^{58}$Ni(P,EL)$^{58}$Ni}\\
\hline
E(MeV) & KD & HF & NM & E(MeV) & KD & HF & NM \\
\hline
11 & 0.7137 & 1.1819 & 0.5797 & 20.4 & 0.2436 & 0.3441 & 0.5361 \\
\hline
26 & 0.8428 & 0.5314 & 0.6489 & 40 & 0.4012 & 0.6274 & 0.7128 \\
\hline
\multicolumn{4}{|c|}{$^{90}$Zr(N,EL)$^{90}$Zr} & \multicolumn{4}{c|}{$^{90}$Zr(P,EL)$^{90}$Zr}\\
\hline
E(MeV) & KD & HF & NM & E(MeV) & KD & HF & NM \\
\hline
11 & 0.1456 & 0.3662 & 0.4944 & 25.05 & 0.1038 & 0.5950 & 0.9651 \\
\hline
24 & 0.2423 & 0.6302 & 0.6855 & 49.35 & 0.4301 & 1.0429 & 1.2387 \\
\hline
\multicolumn{4}{|c|}{$^{208}$Pb(N,EL)$^{208}$Pb} & \multicolumn{4}{c|}{$^{208}$Pb(P,EL)$^{208}$Pb}\\
\hline
E(MeV) & KD & HF & NM & E(MeV) & KD & HF & NM \\
\hline
20 & 0.2819 & 0.5742 & 0.6649 & 24.1 & 0.3585 & 1.0882 & 1.3861 \\
\hline
24 & 0.3440 & 0.5638 & 0.6446 & 30.3 & 0.4304 & 1.5898 & 2.0557 \\
\hline
\end{tabular}
\end{center}
\label{table2}
\caption{Relative {\it rms} deviation of the theoretical differential cross sections with  respect to the experimental values corresponding to the analyzed reactions computed with the Gogny MOP, in both HF and NM prescriptions, and by the POPKD.}  
\end{table}

In nuclear reactions a usual way to assess the quality of a theoretical fit to experimental data is the $\chi^2/N$ test. However, in this work we are more interested in the comparison of the predictions provided by different models and we will use as quality indicator the relative {\it rms} deviation of the theoretical DCS predictions with respect to the corresponding experimental data. This number (multiplied by 100) corresponds to the percentage of the average deviation between the theoretical prediction with respect to the experimental value for a given reaction. Table 2 collects these relative {\it rms} for the reactions considered in the previous subsections. We see that the global POPKD shows a relative {\it rms} ranging from 20\% to 45\%; with some particular reactions this percentage is smaller or larger. The {\it rms} deviations  computed with the Gogny MOP model are larger, and in general, lie between 35\% and 65\% if the HF prescription is used and between 50\% and 85\% when the MOP is calculated with the NM approach. There are few cases where globally the Gogny MOP fails, as for example the elastic proton scattering off $^{208}$Pb as it can be seen from this Table. This result confirms the wrong behavior of the differential cross sections for these reactions displayed in the two lower panels of Figure \ref{fig8} alluring to the failure of the model in these reactions. \\  

In Table 3 we show the total cross sections (neutrons) and reaction cross sections (protons) for the nucleon-nucleus reactions discussed in the previous subsections. We compare the predictions of the Gogny MOP with the results provided by the POPKD. From this Table we see that the total cross sections for neutrons computed with the Gogny MOP roughly follow the predictions of the POPKD calculations. In general, for low incident energies, the Gogny MOP, in both NM and HF approaches, underestimates the phenomenological predictions, while just the contrary happens at higher incident energy. Notice that the total cross section contains the elastic contribution. If in the case of n + $^{208}$Pb reactions the elastic contribution is removed, the reaction cross section computed with the Gogny MOP, for both HF and NM approaches, is larger than the one obtained with the POPKD. The same trend is found for the reaction cross sections in elastic proton scattering. Again at low bombarding energy these cross sections computed with the Gogny MOP underestimates the POPKD values and the contrary happens at higher energies. The differences between the total and reaction cross sections computed with the Gogny MOP and the POPKD predictions are usually less than 10\%, although in some particular cases the discrepancy can increase up to about  15\%. From this analysis of the cross sections provided by the Gogny forces, it seems clear that this model overestimates the absorption when the incident energy is high enough.\\   

\begin{table}
\begin{center}
\begin{tabular}{|c|c|c|c|c|c|c|c|}
\hline
\multicolumn{4}{|c|}{$^{56}$Fe(N,EL)$^{56}$Fe} & \multicolumn{4}{c|}{$^{56}$Fe(P,EL)$^{56}$Fe} \\
\hline
& \multicolumn{3}{c|}{$\sigma_T$(millibarns)} & & \multicolumn{3}{c|}{$\sigma_R$(millibarns)} \\
\hline
E(MeV) & KD & HF & NM & E(MeV) & KD & HF & NM \\
\hline
11 & 3000 & 2723.5 & 2817 & 17.5 & 1069.8 & 1048.9 & 1018.6 \\
\hline
14.7 & 2588 & 2497.0 & 2614.8 & 20.4 & 1091.7 & 1106.7 & 1077.0 \\
\hline
20 & 2322.2 & 2435.6 & 2426.7 & 24.6 & 1102.7 & 1162.5 & 1118.9 \\
\hline
26 & 2331 & 2458.0 & 2321.9 & 35.2 & 1062.6 & 1218.7 & 1169.1 \\
\hline
\multicolumn{4}{|c|}{$^{98}$Mo(N,EL)$^{98}$Mo} & \multicolumn{4}{c|}{$^{58}$Ni(P,EL)$^{58}$Ni}\\
\hline
& \multicolumn{3}{c|}{$\sigma_T$(millibarns)} & & \multicolumn{3}{c|}{$\sigma_R$(millibarns)} \\
\hline
E(MeV) & KD & HF & NM & E(MeV) & KD & HF & NM \\
\hline
11 & 4297.3 & 4124.4 & 3606.3 & 20.4 & 1079.8 & 1060.7 & 1036.9 \\
\hline
26 & 3011.1 & 3222.9 & 3241.9 & 40 & 1032.8 & 1202.0 & 1158.3 \\
\hline
\multicolumn{4}{|c|}{$^{90}$Zr(N,EL)$^{90}$Zr} & \multicolumn{4}{c|}{$^{90}$Zr(P,EL)$^{90}$Zr}\\
\hline
& \multicolumn{3}{c|}{$\sigma_T$(millibarns)} & & \multicolumn{3}{c|}{$\sigma_R$(millibarns)} \\
\hline
E(MeV) & KD & HF & NM & E(MeV) & KD & HF & NM \\
\hline
11 & 4153.8 & 3876.0 & 3518.9 & 25.05 & 1298.6 & 1312.3 & 1254.6 \\
\hline
24 & 2961.3 & 3137.6 & 3145.9 & 49.35 & 1277.6 & 1492.8 & 1426.8 \\
\hline
\multicolumn{4}{|c|}{$^{208}$Pb(N,EL)$^{208}$Pb} & \multicolumn{4}{c|}{$^{208}$Pb(P,EL)$^{208}$Pb}\\
\hline
& \multicolumn{3}{c|}{$\sigma_T$(millibarns)} & & \multicolumn{3}{c|}{$\sigma_R$(millibarns)} \\
\hline
E(MeV) & KD & HF & NM & E(MeV) & KD & HF & NM \\
\hline
20 & 5832.5 & 5443.4 & 4937.1 & 24.1 & 1558.6 & 1451.9 & 1498.0 \\
\hline
24 & 5658.9 & 5195.1 & 4933.1 & 30.3 & 1779.9 & 1818.5 & 1767.3 \\
\hline
\end{tabular}
\end{center}
\label{table3}
\caption{Total and reaction cross sections (in mb) for the elastic neutron and proton scattering on the different targets analyzed in this work.} 
\end{table}

Other relevant quantities to assess the quality of the optical potential are the volume integrals of the real and imaginary contributions to its central part as well as the corresponding {\it rms} radii. These integrals provide a measure of the energy dependence of the optical potential and determine to a large extent the scattering cross sections. These integrals and their {\it rms} radii are defined as 

\begin{eqnarray}
J_v=-\frac{1}{A_T}\int V(r)d\vec{r}, & J_w=-\frac{1}{A_T}\int W(r)d\vec{r}, \\
R_v=\left[\frac{\int V(r)r^2d\vec{r}}{\int V(r)d\vec{r}}\right]^{1/2}, & R_w=\left[\frac{\int W(r)r^2d\vec{r}}{\int W(r)d\vec{r}}\right]^{1/2},
\label{intvol}
\end{eqnarray}

where $A_T$ is the mass number of the target nucleus.

\begin{figure}
\centering
\includegraphics[width=\linewidth,clip=true]{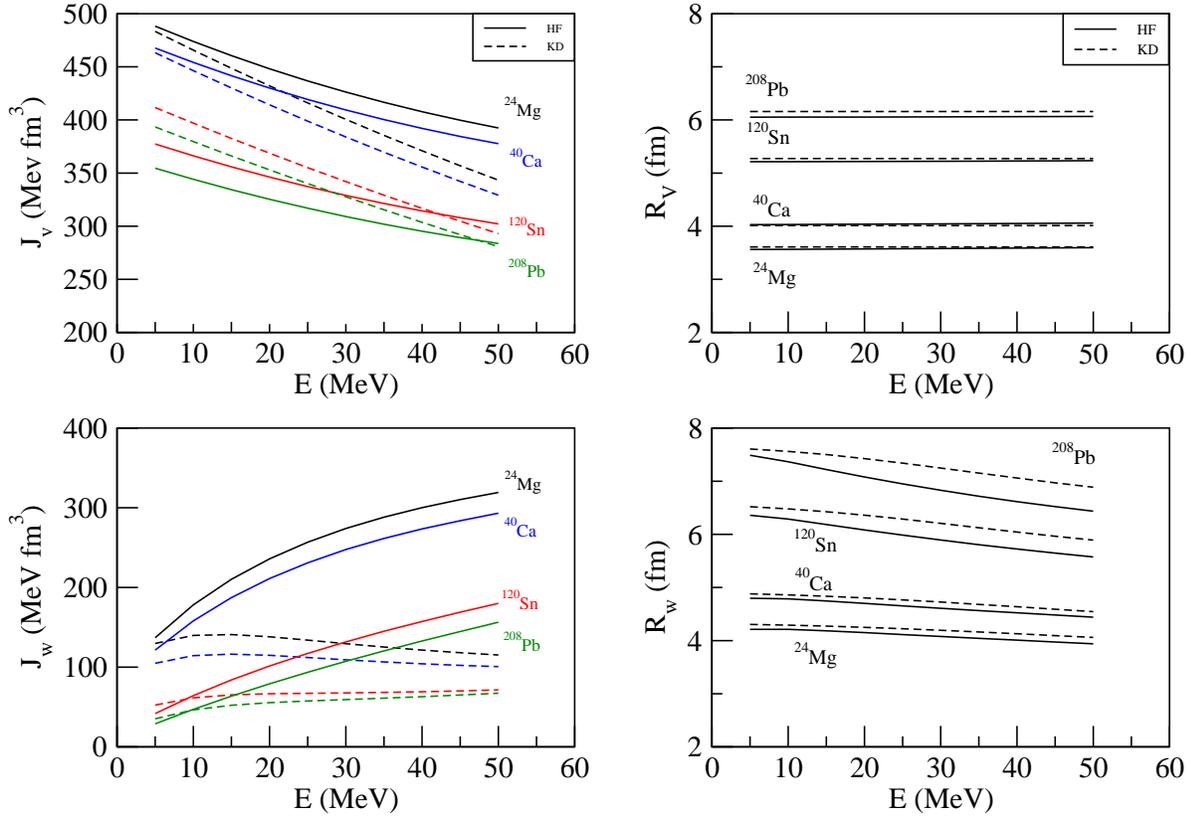}
\setlength\mathindent{70pt}
\caption{Volume integrals and {\it rms} radii of the neutron optical potential calculated with the Gogny HF approach (solid line) and the POPKD (dashed line) as a function of the incident  energy}
\label{fig14}
\end{figure}

The integrals of the real $J_v$ and imaginary $J_w$ central part of the neutron optical potential computed with the Gogny MOP in the HF approach and with the POPKD for several target nuclei from $^{24}$Mg to $^{208}$Pb are displayed as a function of the energy up to 60 MeV in Figure \ref{fig14}. Regarding the integrals of the real contribution, $J_v$ decreases with increasing energy in both POPKD and Gogny MOP calculations. However, $J_v$ decreases faster in  the phenomenological POPKD calculation than those in the Gogny MOP HF. For a given energy $J_v$ also decreases with increasing mass number of the target nucleus. We also see that for asymmetric targets ($^{120}$Sn and $^{208}$Pb) the volume integrals $J_v$ obtained with the Gogny MOP HF approach are also shifted to smaller values than the corresponding POPKD values. The integrals of the imaginary part of the central potential calculated with the POPKD show a  rather constant trend with the energy and also decrease when the mass number of the target nucleus increases. The integrals $J_w$ calculated with the Gogny MOP HF approach show a very different pattern. They increase roughly linearly with the incident energy, in particular for the heavier targets. At low energies the imaginary parts of the POPKD and Gogny MOP HF are strongly peaked at the surface (see middle panels of Figures \ref{fig2} and \ref{fig3}), pointing out that the surface contribution is larger than  the volume one. When the energy increases, this tendency is reversed in such a way that the sum of both contributions remains roughly constant in the range of energies analyzed in this work in the POPKD case. However, in the Gogny MOP HF the imaginary potential is stronger than the one predicted by the POPKD and  shows a larger volume/surface ratio, which increases with the energy more than in the POPKD case, as  can be seen in Figure 2. These differences may explain the different behavior displayed by  $J_w$ in the POPKD and Gogny MOP HF calculations. The {\it rms} of the volume integrals $J_v$ and $J_w$ are almost identical for $R_v$ and show more differences for $R_w$, in particular for the heaviest targets. From this analysis we can see that the energy dependence of the Gogny MOP shows different trend from that  exhibited by the POPKD. To remedy this deficiency of the Gogny MOP, the results displayed in Figure \ref{fig14} suggest that the introduction of energy-dependent renormalization factors, as was done in Ref.\cite{bauge98}, could help to find a better agreement with the behavior exhibited by the POPKD.

\subsection{Nuclear Matter approach to the Microscopic Optical Potential versus Nuclear Structure Model}

\begin{figure}[ht]
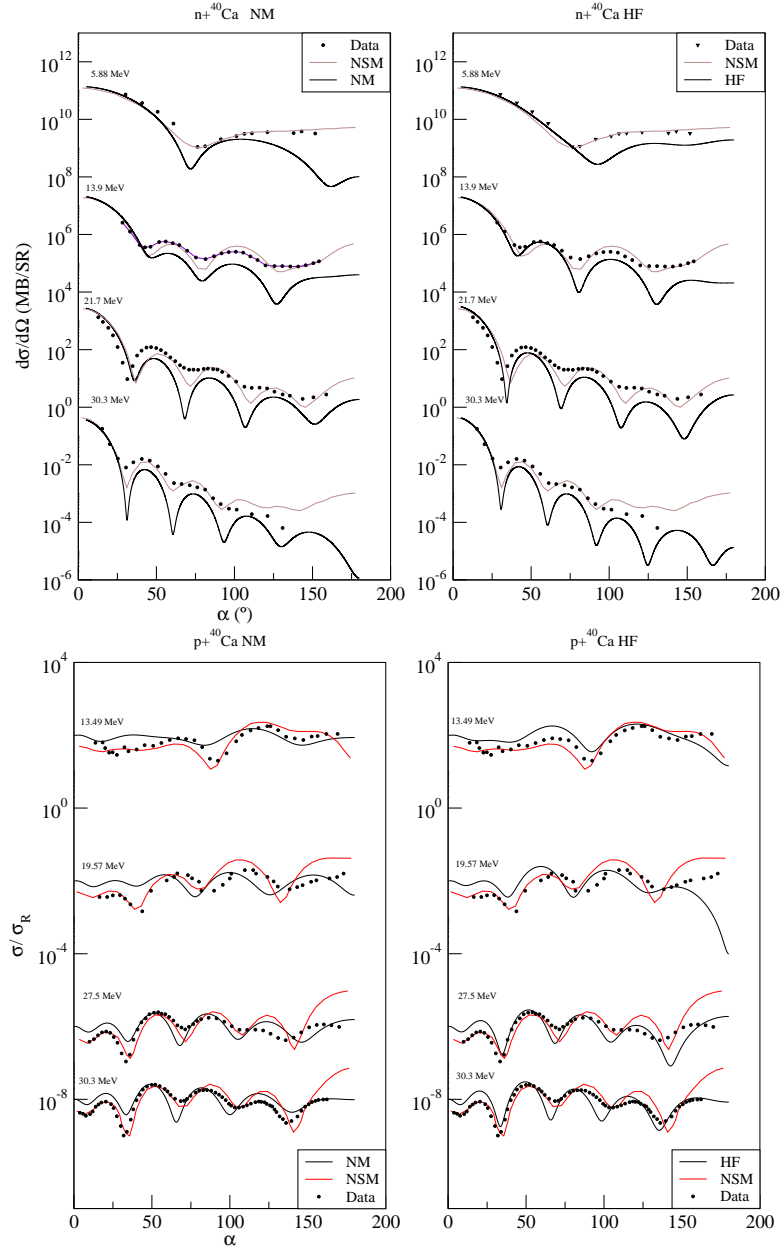

\centering
\includegraphics[width=0.65\linewidth,clip=true]{qtgraceArellanoCa40Neutrons-NM-HF.eps}
\includegraphics[width=0.65\linewidth,clip=true]{qtgraceArellanoCa40Protons-NM-HF.eps}
\setlength\mathindent{80pt}
\caption{Differential cross sections for neutron and proton scattering off $^{40}Ca$}
\label{fig11}
\end{figure}

In this subsection we want to compare the predictions of the Gogny MOP developed in this work with the results obtained using the NSM for describing the neutron and proton elastic scattering by targets of $^{40}$Ca and $^{48}$Ca at several energies using  the D1S interaction, which are reported in Refs.\cite{blanchon15a,blanchon15b,blanchon17}. We refer the reader to these references for technical details about the NSM calculations that have been performed. To this  end, we display in Figure \ref{fig11} the DCS for elastic scattering of neutrons (upper panel)  and protons (lower panel) on $^{40}$Ca at several energies between 5.88 and 30.3 MeV (neutrons) and 13.49 and 30.3 MeV (protons). In this Figure we show the results obtained with our approach, using both the NM (left columns) and HF (right columns) prescriptions, without including the compound nucleus corrections. In the same Figure we also display the NSM predictions and the experimental data both  extracted from \cite{blanchon15b}. In Figure  \ref{fig12} we show the same results but for a target $^{48}$Ca and projectile energies between 4.7 and 16.3 MeV (neutrons) and between 8 and 30 MeV (protons). From these Figures we can see that for the considered reactions, the DCS predicted by our model reproduce fairly well the experimental data up to scattering angles of about 30$^o$. We also see that the DCS patterns predicted by the Gogny MOP are similar to those obtained using the NSM. For neutron scattering our model predicts, using both the NM and HF prescriptions, the position of the dips of the DCS in qualitative good agreement with the results obtained from the NSM calculations, although a shift in the position of the dips can be observed for bombarding energies below 10 MeV for both $^{40}$Ca and $^{48}$Ca targets. Our model also predicts stronger absorption then the NSM calculations as can be  clearly seen in the upper panels of Figures \ref{fig11} and \ref{fig12}. The experimental DCS for proton scattering on $^{40}$Ca are qualitatively well reproduced by the Gogny MOP, which are also in a reasonably good agreement with the predictions of the NSM. However, in the case of a target of $^{48}$Ca there are more discrepancies between our results and the experimental data, although the oscillatory structure of $\sigma/\sigma_R$ is qualitatively reproduced by the Gogny MOP, in particular using the NM prescription. Our results clearly differ from the DCS predicted by the NSM for high proton energies and scattering angles above 50$^{o}$. At relatively high energies and large scattering angles, $\sigma/\sigma_R$ are not well described by the NSM model as stated in \cite{blanchon15b}. It should be pointed out that our model, as well as the NSM of \cite{blanchon15b}, have no free parameters fitted to reproduce experimental scattering data and, therefore, they are fully predictive regarding scattering results, owing to the fact that the Gogny forces of the D1 family do not include scattering data in its fitting protocol. It is also important to note that in spite of the simplicity of our model which does not take into account explicitly the structure of the excited levels in the target nucleus (which is considered on the average in the nuclear matter limit), it is able to describe qualitatively the trends found in the more elaborate NSM calculations.\\ 

\setlength\mathindent{20pt}
\begin{figure}[ht]
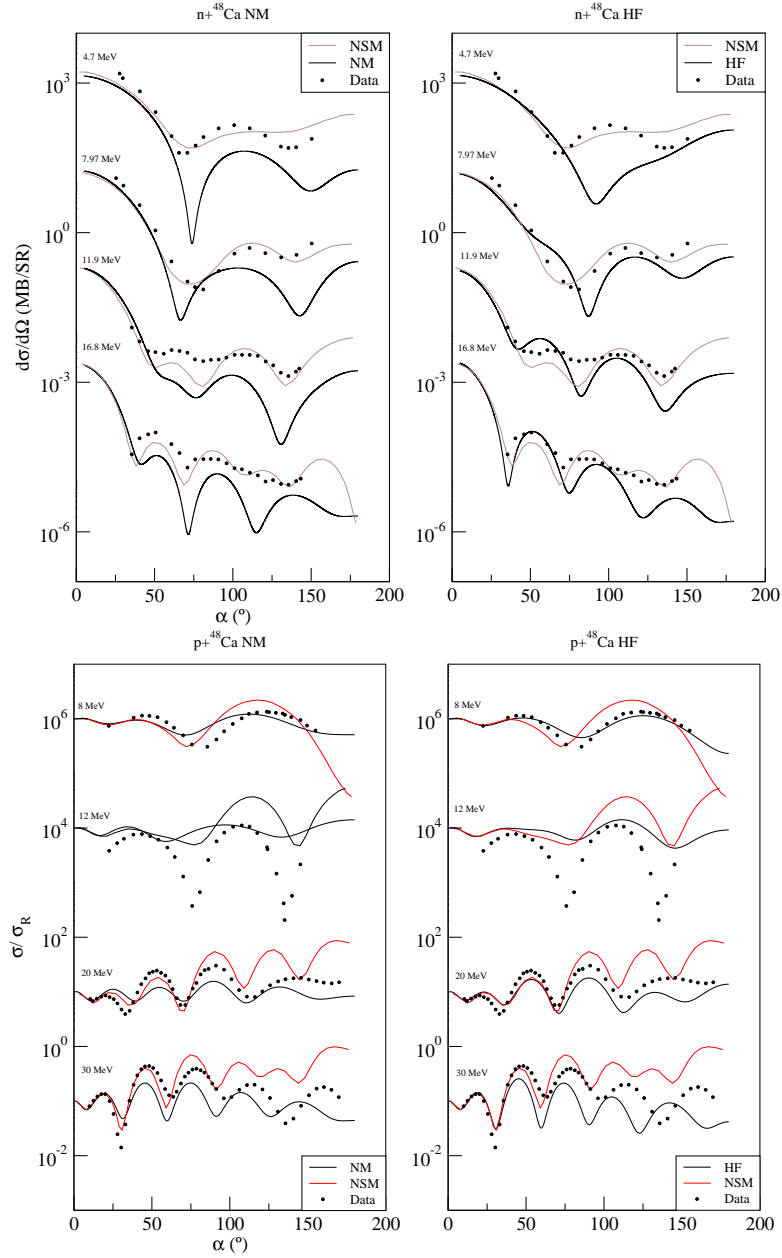

\centering
\includegraphics[width=0.65\linewidth,clip=true]{qtgraceArellanoCa48Neutrons-NM-HF.eps}
\includegraphics[width=0.65\linewidth,clip=true]{qtgraceArellanoCa48Protons-NM-HF.eps}
\setlength\mathindent{80pt}
\caption{Differential cross sections for neutron and proton scattering off $^{48}Ca$}
\label{fig12}
\end{figure}
\setlength\mathindent{20pt}

To be more precise about the quality of the theoretical calculations of the differential cross sections discussed before and displayed in Figures \ref{fig11} and \ref{fig12}, we report in Table 4 the corresponding relative {\it rms} deviation between the experimental and the theoretical values obtained through the POPKD, Gogny MOP and the NSM. From this Table we can see that the NSM predicts the more accurate description of the experimental data for neutron scattering with energies below of 20 MeV. However, this model shows some deficiencies above this energy. For example, the relative {\it rms} deviation for the reaction n+$^{40}$Ca at energy of 30.3 MeV for the incident projectile is close to unity, which is consistent with the large deviation of the NSM differential cross sections for backwards angles displayed in Figure \ref{fig11}. The POPKD predicts, for the considered energies, {\it rms} deviations between 10\% and 35\%, which in the Gogny MOP model are large, between 25\%-65\% in the HF approach and 25\%-60\% using the NM prescription.

\begin{table}
\begin{center}
\begin{tabular}{|c|c|c|c|c|}
\hline
\multicolumn{5}{|c|}{$^{40}$Ca(N,EL)$^{40}$Ca}\\
\hline
E (MeV) & KD & HF & NM & NSM  \\
\hline
5.88 & 0.3445 & 0.3617 & 0.2494 & 0.0669 \\
\hline
13.9 & 0.1700 & 0.4921 & 0.4243 & 0.1378 \\
\hline
21.7 & 0.1729 & 0.4754 & 0.4899 & 0.3385  \\
\hline
30.3 & 0.1269 & 0.6508 & 0.5882 & 0.9454  \\
\hline
\end{tabular}
\end{center}

\begin{center}
\begin{tabular}{|c|c|c|c|c|}
\hline
\multicolumn{5}{|c|}{$^{48}$Ca(N,EL)$^{48}$Ca}\\
\hline
E (MeV) & KD & HF & NM & NSM  \\
\hline
4.7  & 0.3162 & 0.2938 & 0.3885 & 0.1015 \\
\hline
7.97 & 0.3194 & 0.3114 & 0.3286 & 0.1049 \\
\hline
11.9 & 0.2387 & 0.2846 & 0.5146 & 0.1582  \\
\hline
16.8 & 0.1449 & 0.2476 & 0.4188 & 0.1342  \\
\hline
\end{tabular}
\end{center}
\label{table4}
\caption{Relative {\it rms} deviation of the theoretical differential cross sections respect to the experimental values corresponding to the n + $^{40,48}$Ca computed with the Gogny MOP (HF and NM prescriptions), the POPKD and the NSM.}
\end{table}

In the recent years, there is a growing interest in performing {\it ab initio} calculations of the MOP on the basis of the effective field theory. These calculations, which include some single-particle states to obtain the imaginary part, have been mainly applied to study neutron scattering on light-medium nuclei, such as $^{16}$O \cite{idini19} and $^{40,48}$Ca \cite{rotureau18,idini19,whitehead20}, where the nuclear structure can be obtained in a relatively simple way. On the other hand, MOPs based on effective forces of Skyrme and Gogny types where the NM approach is used to compute the imaginary part, are more suitable to be applied to any nucleus, because in this case the nuclear structure of each nucleus is only taken into account on average. The differential cross sections obtained with {\it ab initio} MOP reported in the literature \cite{rotureau18,idini19,whitehead19,whitehead20} reproduce fairly well the experimental data with a quality similar to that found in the NSM or Gogny MOP calculations. The agreement of the MOPs computed with chiral forces with the experiment as well as with the predictions of the POPKD can be improved if the imaginary part is fine tuned to incorporate, in an effective way, additional contributions not accounted in the {\it ab initio} calculations. \\ 

\begin{figure}
\centering
\includegraphics[width=0.60\linewidth,clip=true]{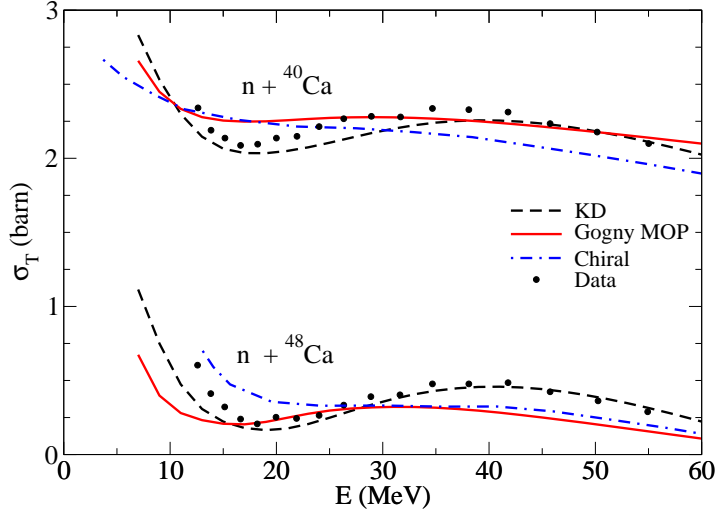}
\setlength\mathindent{20pt}
\caption{The neutron total cross section for n+$^{40,48}$Ca. The curves corresponding to  $^{48}$Ca are shifted down by 2 barns. The results obtained withh the POPKD and the values reported in Ref.\cite{whitehead20} are also displayed.}
\label{fig15}
\end{figure}

As an example and in order to be more quantitative, we display in Figure \ref{fig15} the total cross section for the n+$^{40,48}$Ca reactions computed with the aforementioned models as a function of the energy of the incident neutrons up to 60 MeV and  compare also with the experimental data taken from Ref.\cite{shane10}. In the same Figure we also display the very recent results extracted from Figure 10 of the Ref.\cite{whitehead20}, which are obtained within the JLM scheme with a chiral interaction. The total cross sections calculated with the POPKD reproduce well the behavior of the experimental data. The total cross sections computed with the Gogny MOP in the HF approach follow the average trend of the experimental values. For the n+$^{40}$Ca reaction, the Gogny MOP total cross section overestimates the experimental values for incident energies between 10 and 25 MeV and from this energy on follows rather accurately the experimental data. For the  n+$^{48}$Ca reaction, the Gogny MOP reproduces the experimental total cross sections from about 15 MeV till 30 MeV. For higher energies, our model underestimates the experimental as well the POPKD results. The chiral MOP calculation of Ref.\cite{whitehead20} predicts in the considered range of energies a decreasing trend, overestimating the experimental values between 10 and 30 MeV and underestimating them above this energy.  

\section{Summary and outlook}

 We have developed a relatively simplified microscopic optical potential suitable to estimate scattering properties predicted by finite-range effective nuclear interactions. The numerical applications in this work have been carried out for the Gogny D1S force. It is worthwhile to point out that the theoretical model derived in this work could also be easily applied to other finite-range interactions, such as the Momentum Dependent Interactions (MDI) \cite{das03,chen14} or the Simple Effective Interactions (SEI) \cite{behera05,behera13}. Inspired by the optical model calculation performed by Jeukene, Lejeune and Mahaux in nuclear matter, we determine the real and imaginary parts of the central potential from the first and second terms of the Taylor expansion of the mass operator in nuclear matter. The real part of the central potential in our model is calculated as the single-particle potential felt by the projectile due to the nucleons of the target in two different prescriptions. In one of them, we strictly follow the nuclear matter approach and compute the real part as the Hartree-Fock potential in the infinite medium considering only the two-body part of the effective nucleon-nucleon interaction (NM approach) as has been done in Ref.\cite{shen09}. In the other prescription the real part is provided by the Hartree-Fock potential of the target at Slater level, which takes explicitly into account the finite size of the nucleus in the direct term and the rearrangement contribution coming from the density-dependent part of the force (HF approach), as has been done in some earlier works \cite{pilipenko10,pilipenko12}. An important approximation, needed to simplify the calculation of the imaginary part, consists of the parabolic expansion of the momentum dependent exchange part of the single-particle potential around the Fermi momentum, which explicitly sorts out the effective mass hidden in the full exchange term. This microscopic optical potential is applied to finite nuclei through the Local Density Approximation using the neutron and proton densities of the target nucleus computed self-consistently  with the same interaction at Hartree-Fock level within the quasi-local energy density theory.  Notice that this microscopic optical potential is fully predictive, as far as there are no free parameters to be fitted to nucleon-nucleus scattering data. Therefore, this model allows to calibrate the ability of  a given interaction for describing elastic nucleon-nucleus reactions. It is found that the real and imaginary central parts of our microscopical optical potential follow reasonably well the trends of the phenomenological potential of Koning and Delaroche \cite{koning03}, in particular at low energies. We compare the differential cross sections and analyzing powers calculated with our microscopical optical potential for different neutron and proton elastic reactions on different targets at several energies with the results obtained with the potential of Koning and Delaroche as well as with the experimental data. Our relatively simple model is able to reproduce quite accurately the experimental differential cross sections and analyzing powers for forward scattering angles up to 30$^o$-50$^o$ for neutron and proton elastic scattering on targets between $^{40}$Ca and $^{208}$Pb in an energy range of the projectile between 10-30 MeV. For larger scattering angles the differential cross sections computed with our model may differ more from  the experimental data, in particular for neutrons, where our results underestimate the experimental data. The pattern of the differential cross sections and analyzing powers predicted by the phenomenological potential of Koning and Delaroche is also qualitatively well reproduced by our model, although the position of the dips is shifted with respect to those corresponding to the phenomenological potential in some cases. Notice that at high energies the imaginary part predicted by our model is stronger than that in the phenomenological potential. This implies smaller differential cross sections and more damped oscillations in the analyzing power as compared  to those from the phenomenological potential. To assess the global quality of the predictions of our Microscopical Optical Potential based on the Gogny interaction, we have computed the relative {\it rms} deviation of the theoretical differential cross sections with  respect to the corresponding experimental values. For each reaction considered in this work, the value of this indicator is compared with the one provided by the phenomenological model of Koning and Delaroche, used here as a benchmark. To have more global insight about the predictive power of our model, we have also analyzed the total and reaction cross sections, for neutrons and protons, respectively. This study shows that our model overestimates the absorption, specially at high energy. This points out some deficiencies in the energy dependence predicted by our model compared to the corresponding behavior obtained with the phenomenological potential of Koning and Delaroche. This fact is analyzed in more detail by computing the volume integrals of the real and imaginary parts of the central contribution to the neutron optical potential based on the Gogny forces. Globally, we find that, as a function of the energy of the incident projectile, the real part roughly follows the trends of the volume integrals of the model of Koning and Delaroche, although with slightly different slope. However, the volume integrals of the imaginary contribution computed with our Gogny based optical potential clearly differ from the predictions of the phenomenological potential, which may be a possible reason for the strong absorption exhibited by our model, which in turn is the reason of the lack of accuracy in the predictions of our model for large scattering angles. \\    

We have also compared  the differential cross sections of neutron and proton elastic scattering on targets of $^{40}$Ca and $^{48}$Ca computed with our microscopic optical model with the results reported by Blanchon and collaborators obtained by using the more fundamental nuclear structure model. In spite of the fact that we have not included the compound nucleus contribution, our predicted differential cross sections for forward angles up to about 30$^o$-50$^o$ agree rather well with the Blanchon's results for all the considered energies. However, for larger scattering angles our model predicts smaller differential cross sections than the nuclear structure model. In the case of proton scattering our calculation reproduces qualitatively the experimental differential cross sections in the whole range of measured data. Our results are also in good agreement with the nuclear structure model predictions for forward angles and show more differences at large backwards angles. We have also compared the total cross section predicted by our models as a function of the energy with the experimental data, with the values provided by the phenomenological model of Koning and Delaroche and with the recent results of Ref.\cite{whitehead20} computed using an optical model based on chiral forces. We find that the experimental data are well reproduced by the phenomenological model of Koning and Delaroche, while our model as well as the one of Ref.\cite{whitehead20} only reproduce the average trend. Although we have not reported the results here, we have checked that our model used with another Gogny forces such as D1N or D1M, describes the nucleon-nucleus elastic reactions with a similar quality to that reported in the present work with the D1S interaction.\\ 

Some possible extensions of this work are the following. The theoretical description of the Hartree-Fock potential generated by the nucleons of the target interacting with the projectile, which in the present work is computed at Slater level, may be improved by including $\hbar^2$ corrections to the exchange part of the potential. We have seen that although the Gogny D1S interaction can give a qualitative description of the neutron and proton elastic scattering at moderate energies of the projectile, the energy dependence of the imaginary part shows some deficiencies. To cure this problem, renormalization factors of the real and imaginary parts of the microscopic optical potential computed in this work, which may depend on the energy of the projectile and on the mass number of the target, are needed to improve the agreement with the experimental data. In this way one can obtain a semi-microscopic nucleon-nucleus optical potential based on the Gogny interactions which is able to describe neutron and proton elastic scattering in a wide range of mass numbers of the target and energies of the projectile with a quality similar to that found using phenomenological models. Our model could also be generalized in a rather straightforward manner to describe the elastic scattering of light projectiles, such as $^2$H, $^3$H, $^3$He and $^4$He, in a similar way that has been done by Shen Quing Biao and collaborators in Refs. \cite{guo10,guo14,guo09,guo11}. Another possible extension of our work is to study  quasi-elastic $(p,n)$ reactions that allow to study the isoscalar and isovector parts of the optical potential in the spirit of the Lane model, which in turn are connected to the symmetry energy of the underlying nucleon-nucleon effective interaction. 

\section*{Acknowledgement}

The authors acknowledge support from Grant CEX2019-000918-M from State Agency for Research of the Spanish Ministry of Science and Innovation through the "Unit of Excellence Martia de Maezu 2020-2023" award to ICCUB. X.V. also acknowledges support from Grant FIS2017-87534-P from MINECO and FEDER. The authors are indebted to P. Schuck, J.N. De, Tapas Sil, C. Mondal and C. Gonzalez-Boquera for useful discussions.\\

\section{Appendix 1}

To evaluate the integral

\begin{equation}
I_i=\int d\vec{k}_\nu d\vec{k}_\lambda d\vec{k}_\mu f_i(\gamma_1,\gamma_2) \delta(\vec{k}_\alpha+\vec{k}_\nu-\vec{k}_\lambda-\vec{k}_\mu)\delta(\beta_{\tau_\alpha} k_\alpha^2+\beta_{\tau_\nu} k_\nu^2-\beta_{\tau_\alpha} k_\lambda^2-\beta_{\tau_\nu} k_\mu^2), \nonumber \\ 
\label{eqA1}
\end{equation}

we start performing a change of variables passing from $\vec{k}_\lambda$ and $\vec{k}_\mu$ to the corresponding center of mass and relative coordinates: 

\[\vec{k}_p=\vec{k}_\lambda+\vec{k}_\mu=\vec{k}_\alpha+\vec{k}_\nu,\qquad \vec{q}_p=\frac{1}{2}(\vec{k}_\lambda-\vec{k}_\mu),\]

which allows to write \cite{shen09} 

\[\vec{k}_\lambda=\frac{\vec{k}_p}{2}+\vec{q}_p,\qquad \vec{k}_\mu=\frac{\vec{k}_p}{2}-\vec{q}_p.\]

\begin{figure}
\centering
\includegraphics[width=0.4\linewidth]{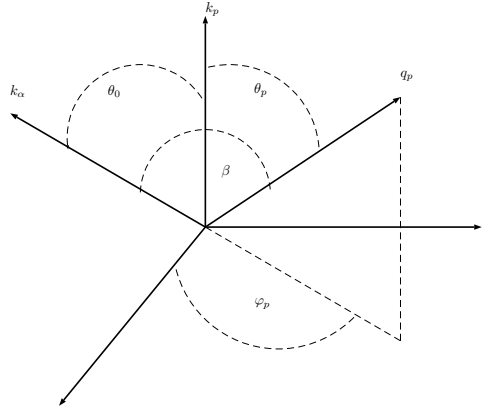}
\setlength\mathindent{160pt}
\caption{Reference system}
\label{fig13}
\end{figure}

According to Figure \ref{fig13}, we define the following angles and vectors 

\begin{eqnarray}
\mu\equiv\cos(\theta_\nu)=\cos(\vec{k}_\alpha,\vec{k}_\nu),  & \qquad \mu_p\equiv\cos(\theta_p)\equiv\cos(\vec{k}_p,\vec{q}_p), \nonumber \\ \cos(\theta_0)\equiv\cos(\vec{k}_\alpha,\vec{k}_p), & \qquad \cos(\beta)\equiv\cos(\vec{k}_\alpha,\vec{q}_p), \nonumber \\ \hat{k}_\alpha=\left(\sin(\theta_0),0,\cos(\theta_0)\right), \nonumber & \qquad \hat{q}_p=\left(\sin(\theta_p)\cos(\varphi_p),\sin(\theta_p)\sin(\varphi_p),\cos(\theta_p)\right), \nonumber
\end{eqnarray}

from where we can write

\[\cos(\theta_0)=\frac{k_\alpha+k_\nu\mu}{k_p},\]

and

\[\cos(\beta)=\sin(\theta_0)\sin(\theta_p)\cos(\varphi_p)+cos(\theta_0)\cos(\theta_p).\]

\setlength\mathindent{20pt}

Defining now the following constants:

\[A=\frac{\gamma_1^2}{4}+\frac{\gamma_2^2}{4},\qquad B=\frac{\gamma_2^2}{4}-\frac{\gamma_1^2}{4},\]

the functions $f_i$ defined in the main text can be recast as:

\[f_i(\gamma_1,\gamma_2)=e^{-\frac{\gamma_1^2}{4}k_{\alpha\lambda}^2-\frac{\gamma_2^2}{4}k_{\alpha\mu}^2}=\]

\[e^{-\frac{A}{4}k_\alpha^2-\frac{A}{4}k_\nu^2+\frac{A}{2}k_\alpha k_\nu\mu-Aq_p^2-B\frac{q_p}{k_p}[k_\alpha^2-k_\nu^2]\cos(\theta_p)-2Bk_\alpha q_p\sin(\theta_0)\sin(\theta_p)\cos(\varphi_p)}.\]

Introducing a new variable $C$ defined as

\[C=2Bk_\alpha q_p\sin(\theta_0)\sin(\theta_p),\]

and taking into account the integral

\[\int_0^{2\pi}e^{-c\cos(\varphi_p)}d\varphi_p=2\pi I_0(C), \]

where $I_0(x)$ is the zeroth-order modified Bessel function, we can finally write the integral $I_i$ in the following way:

\begin{equation}
I_i=\frac{(2\pi)^2}{\beta_{\tau\alpha}-\beta_{\tau\nu}}e^{-\frac{A}{4}k_\alpha^2}\int_{\sqrt{B_0}}^{k_{F\nu}}dk_\nu k_\nu^2e^{-\frac{A}{4}k_\nu^2}\int_{-1}^1\frac{d\mu}{k_p}e^{\frac{A}{2}k_\alpha k_\nu\mu}\int_{Q_1}^{Q_2}dq_pq_pe^{-Aq_p^2-B\frac{q_p}{k_p}[k_\alpha^2-k_\nu^2]\cos(\theta_p)}I_0(C), \nonumber \\ \label{eqA2}
\end{equation}

where the limits of the integral over $q_p$ are given by 

\[Q_2^2=\frac{H_2+\beta_{\tau\nu}k_\nu^2-\frac{1}{2}\beta_{\tau\nu}k_p^2}{2\beta_{\tau\nu}},\]

and 

\[Q_1^2=\frac{H_1+\beta_{\tau\nu}k_\nu^2-\frac{1}{2}\beta_{\tau\alpha}k_p^2}{2\beta_{\tau\alpha}},\]

with

\[H_2=\beta_{\tau\alpha}k_\alpha^2-(\beta_{\tau\alpha}-\beta_{\tau\nu})k_{\tau\alpha}^2,\]

and

\[H_1=\beta_{\tau\alpha}k_\alpha^2+(\beta_{\tau\alpha}-\beta_{\tau\nu})k_{\tau\nu}^2,\]

respectively. In general the triple integral (\ref{eqA2}) can not be performed  analytically and therefore we have computed numerically all the integrals $I_1\dots I_{10}$. \\

There is, however, a particular integral, namely $I_6$, where the integral can be obtained analytically. In this case taking into account that

\[f_6=e^{-\frac{\mu_k^2}{4}(k_{\alpha\lambda}^2+k_{\alpha\mu}^2)}=e^{-\frac{\mu_k^2}{8}k_p^2-\frac{\mu_k^2}{2}q_p^2+\frac{\mu_k^2}{2}k_\alpha k_\nu\mu},\] 

we can write this integral as

\begin{equation}
I_6=\frac{(2\pi)^2}{\beta_{\tau\alpha}-\beta_{\tau\nu}}\int_{\sqrt{B_0}}^{k_{F\nu}}k_\nu^2dk_\nu\int_{-1}^1\frac{d\mu}{k_p}e^{\frac{\mu_k^2}{2}k_\alpha k_\nu\mu-\frac{\mu_k^2}{8}k_p^2}\int_{Q_1}^{Q_2}q_pdq_pe^{-\frac{\mu_k^2}{2}q_p^2}, \label{eqA3}
\end{equation}

which can be split as

\[I_6=\left(I_6\right)_1-\left(I_6\right)_2.\]

Performing the integrals over the center of mass momentum $q_p$, and the angle $\mu$ between $\vec{k}_\alpha$ and $\vec{k}_\nu$, one can recast $(I_6)_1$ and $(I_6)_2$ as single integrals over the variable $\vec{k}_\nu$ as  

\begin{eqnarray}
\left(I_6\right)_1&=&\frac{4\pi^2\sqrt{\pi}e^{-\frac{\mu_k^2}{2}k_\alpha^2}e^{-\frac{\mu_k^2}{4}\frac{\beta_{\tau\alpha}-\beta_{\tau\nu}}{\beta_{\tau\alpha}}k_{F\nu}^2}}{\mu_k^3k_\alpha\left(\beta_{\tau\alpha}-\beta_{\tau\nu}\right)}\int_{\sqrt{B_0}}^{k_{F\nu}} k_\nu dk_\nu e^{-\frac{\mu_k^2}{4}\left(1+\frac{\beta_{\tau\nu}}{\beta_{\tau\alpha}}\right)k_\nu^2}\nonumber \\ &&\Big\{erfi\left[\frac{\mu_k}{2}(k_\alpha+k_\nu)\right]-erfi\left[\frac{\mu_k}{2}(k_\alpha-k_\nu)\right]\Big\}, \label{eqA4}
\end{eqnarray}

and

\begin{eqnarray}
\left(I_6\right)_2&=&\frac{4\pi^2\sqrt{\pi}e^{\frac{\mu_k^2}{4}\frac{\beta_{\tau\alpha}-\beta_{\tau\nu}}{\beta_{\tau\nu}}k_{F\alpha}^2}e^{-\frac{\mu_k^2}{4}\left(1+\frac{\beta_{\tau\alpha}}{\beta_{\tau\nu}}\right)k_\alpha^2}}{\mu_k^3 k_\alpha(\beta_{\tau\alpha}-\beta_{\tau\nu})}\int_{\sqrt{B_0}}^{k_{F\nu}} k_\nu dk_\nu e^{-\frac{\mu_k^2}{2}k_\nu^2} \nonumber \\ &&\Big\{erfi\Big[\frac{\mu_k}{2}(k_\alpha+k_\nu)\Big]- erfi\Big[\frac{\mu_k}{2}(k_\alpha-k_\nu)\Big]\Big\} \label{eqA5}.
\end{eqnarray}

The remaining integrals  $(I_6)_1$ and  $(I_6)_2$ can also be performed analytically. The expressions (\ref{eqA4}) and (\ref{eqA5}) correspond to the case of different type of nucleons ($\beta_{\tau\alpha} \ne \beta_{\tau\nu}$). In the case of two identical nucleons, ($\beta_{\tau\alpha}=\beta_{\tau\nu}$), we shall evaluate the limit $\tau_\alpha \to \tau_\nu$ of Eqs.(\ref{eqA4}) and (\ref{eqA5}) obtaining finally

\begin{eqnarray}
I_6&=&\frac{\pi^2\sqrt{\pi}}{\beta_{\tau\alpha} k_\alpha \mu_k}e^{-\frac{\mu_k^2}{2}k_\alpha^2}\int_{\sqrt{B_0}}^{k_{F\nu}} k_\nu\Big[k_\alpha^2+k_\nu^2-2k_{F\alpha}^2\Big]e^{-\frac{\mu_k^2}{2}k_\nu^2} \nonumber \\ &&\Big\{erfi\left[\frac{\mu_k}{2}(k_\alpha+k_\nu)\right]-erfi\left[\frac{\mu_k}{2}\left(k_\alpha-k_\nu\right)\right]\Big\}. \label{eqA6}
\end{eqnarray}

For this particular integral $I_6$ we have checked that the numerical integrals (\ref{eqA2}), with $\beta_\alpha$ and $\beta_\nu$ equal or different, coincide very accurately with the analytical value obtained from (\ref{eqA3}) using Eqs.(\ref{eqA4})-(\ref{eqA6}), which validates our numerical method to obtain the integrals $I_i$. The integral $I_{11}$ has also been calculated exactly in the case of Skyrme forces in Ref.\cite{shen09}. As far as the spin-orbit part of the interaction is the same for Gogny and Skyrme forces, we report here for a sake of completeness the explicit value of the integral $I_{11}$.

\begin{equation}
I_{11}=\int d\vec{k}_\nu d\vec{k}_\lambda d\vec{k}_\mu \left(\vec{k}_{\alpha\nu}\times\vec{k}_{\lambda\mu}\right)^2\delta(\vec{k}_\alpha+\vec{k}_\nu-\vec{k}_\lambda-\vec{k}_\mu) \delta\left(\beta_\alpha k_\alpha^2+\beta_\nu k_\nu^2-\beta_\alpha k_\lambda^2-\beta_\nu k_\mu^2\right), \nonumber \\
\end{equation}

\[\left(\vec{k}_{\alpha\nu}\times\vec{k}_{\lambda\mu}\right)^2=\left|\vec{k}_{\alpha\nu}\right|^2\left|\vec{k}_{\lambda\mu}\right|^2-\left(\vec{k}_{\alpha\nu}\cdot\vec{k}_{\lambda\mu}\right)^2,\]

\begin{equation}
(I_{11})_1=\frac{(2\pi)^2}{\beta_\alpha-\beta_\nu}\int k_\nu^2\left|\vec{k}_{\alpha\nu}\right|^2dk_\nu\int_{-1}^1 \frac{d\mu}{k_p}\int_{Q_1}^{Q_2}q_p\left|\vec{k}_{\lambda\mu}\right|^2dq_p,
\end{equation}

\begin{equation}
(I_{11})_2=\frac{(2\pi)^2}{\beta_\alpha-\beta_\nu}\int k_\nu^2dk_\nu\int_{-1}^1 \frac{d\mu}{k_p}\int_{Q_1}^{Q_2}q_p\left(\vec{k}_{\alpha\nu}\cdot\vec{k}_{\lambda\mu}\right)^2dq_p,
\end{equation}

\[I_{11}=(I_{11})_1-(I_{11})_2,\]

\begin{eqnarray}
(I_{11})_1=\frac{2\pi^2}{945\beta_{\tau\alpha}k_\alpha}\Big\{\big[35B_{16}k_{\tau\nu}^6+45(B_{15}-B_0B_{16})k_{\tau\nu}^4+63(B_{14}-B_0B_{15})k_{\tau\nu}^2-\nonumber \\ 105B_0B_{14}\big]k_{\tau\nu}^3+2(21B_{14}+9B_0B_{15}+5B_0^2B_{16})B_0^{\frac{5}{2}}\Big\},\nonumber
\end{eqnarray}

\begin{eqnarray}
(I_{11})_2=\frac{2\pi^2}{945\beta_{\tau\alpha}k_\alpha}\Big\{\big[35B_{13}k_{\tau\nu}^6+45(B_{12}-B_0B_{13})k_{\tau\nu}^4+63(B_{11}-B_0B_{12})k_{\tau\nu}^2-\nonumber \\ 105B_0B_{11}\big]k_{\tau\nu}^3+2(21B_{11}+9B_0B_{12}+5B_0^2B_{13})B_0^{\frac{5}{2}}\Big\},\nonumber
\end{eqnarray}

\begin{eqnarray}
S&=&\frac{1}{\beta_{\tau\alpha}}(\beta_{\tau\alpha}+\beta_{\tau\nu}),\nonumber \\ F_1&=&\frac{1}{\beta_{\tau\alpha}\beta_{\tau\nu}}\Big[(\beta_{\tau\alpha}+\beta_{\tau\nu})\beta_{\tau\alpha}k_\alpha^2-(\beta_{\tau\alpha}-\beta_{\tau\nu})(\beta_{\tau\alpha}k_{\tau\alpha}^2-\beta_{\tau\nu}k_{\tau\nu}^2)\Big],\nonumber \\ B_0&=&\frac{1}{\beta_{\tau\nu}}\Big[\beta_{\tau\nu}k_{\tau\nu}^2-\beta_{\tau\alpha}(k_\alpha^2-k_{\tau\alpha}^2)\Big],\nonumber \\ B_{14}&=&3(F_1-k_\alpha^2)k_\alpha^2,\nonumber \\ B_{15}&=&5F_1+(3S-2)k_\alpha^2,\nonumber \\ B_{16}&=&5S-\frac{7}{5},\nonumber \\ B_7&=&-\frac{1}{2\beta_{\tau\alpha}}(\beta_{\tau\alpha}-\beta_{\tau\nu}), \nonumber \\ B_6&=&(k_{\tau\alpha}^2-k_{\tau\nu}^2)+\frac{1}{2\beta_{\tau\alpha}}(\beta_{\tau\alpha}-\beta_{\tau\nu})B_0,\nonumber \\ B_{13}&=&3B_7^2+4S+\frac{1}{4}S^2+\frac{4}{5},\nonumber \\ B_{11}&=&3B_6^2+\frac{1}{4}S^2B_0^2,\nonumber \\ B_{12}&=&4F_1-4k_\alpha^2+6B_6B_7-\frac{1}{2}S^2B_0.\nonumber \\
\end{eqnarray}

\section{Appendix 2}

\begin{figure}[hb]
\centering
\includegraphics[width=0.60\linewidth,clip=true]{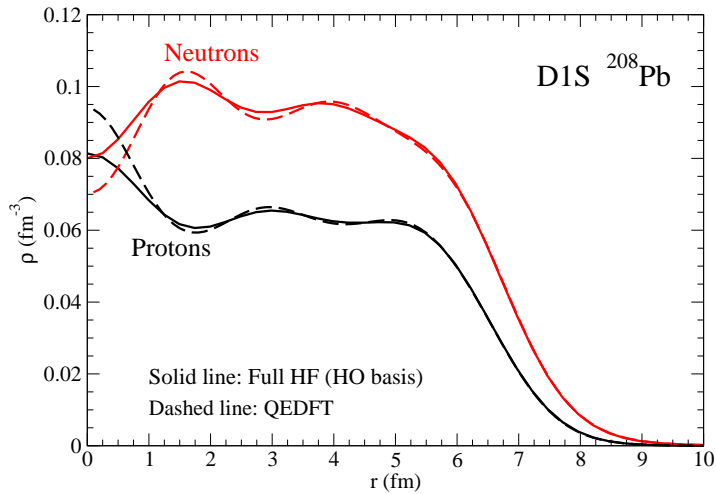}
\setlength\mathindent{20pt}
\caption{Neutron and proton densities of the nucleus $^{208}$Pb computed in the QEDFT approach using the D1S Gogny interaction.}
\label{fig16}
\end{figure}

In this Appendix we briefly outline our method for solving the HF equations to obtain the neutron and proton densities of the target nucleus, which in turn are needed to calculate the MOP based on the Gogny force. The starting point is the so-called quasi-local energy density functional theory (QEDFT), developed for dealing with the HF problem in finite nuclei  using finite-range effective interactions, which was established in Refs.\cite{gridnev98,soubbotin00,soubbotin03} and also discussed in some detail in Refs.\cite{krewald06,behera13,behera16,gonzalez19,mondal19}. The basic idea of the QEDFT applied to finite-range effective forces consists of writing the HF energy density in a local form using a suitable expansion of the HF one-body density matrix. This idea was proposed in the past by Negele and Vautherin \cite{negele72} and also used by Campi and Bouyssy \cite{campi78}. In our calculation we use the density matrix developed in \cite{soubbotin00}, which is based on a Extended Thomas-Fermi (ETF) expansion of the density matrix. This approximation, whose theoretical basis is developed in Refs.\cite{soubbotin03,krewald06}, allows to write the HF energy density in a quasi-local form. To this end we start from the HF energy density, which for a Gogny interaction of the D1 family can be written as 

\begin{eqnarray}\label{eq43}
\mathcal{H}=\mathcal{H}_{kin}+\mathcal{H}_{dir}+\mathcal{H}_{exch}+\mathcal{H}_{zr}+\mathcal{H}_{coul}+\mathcal{H}_{SO},
\end{eqnarray}

where the HF energy density splits into the kinetic and the potential direct, exchange, zero-range, Coulomb and spin-orbit contributions, which are given by:

\begin{eqnarray}\label{eq44}
\mathcal{H}_{kin}&=&\frac{\hbar^2}{2m}(\tau_n+\tau_p), \\ \mathcal{H}_{dir}&=&\frac{1}{2}\sum_i\int d\vec{r^{\prime}} \Bigg\{\left( W_i+\frac{B_i}{2}\right)\rho(\vec{r})\rho(\vec{r^{\prime}})\nonumber \\ &&-\left(H_i+\frac{M_i}{2}\right)\left[\rho_n(\vec{r})\rho_n(\vec{r^{\prime}}) +\rho_p(\vec{r})\rho_p(\vec{r^{\prime}})\right]\Bigg\} e^{-\frac{(\vec{r}-\vec{r^{\prime}})^2}{\mu_i^2}}, \\ \mathcal{H}_{exch}&=&-\frac{1}{2}\sum_{i=1}^2 e^{-\frac{(\vec{r}-\vec{r^{\prime}})^2}	{\mu_i^2}}\Bigg\{\left(B_i+\frac{W_i}{2}\right)\big[\rho_n^2(\vec{r},\vec{r^{\prime}})+\rho_p^2(\vec{r},\vec{r^{\prime}})\big] \nonumber \\ &&-\left(M_i+\frac{H_i}{2}\right)\left(\rho_n(\vec{r},\vec{r^{\prime}})+\rho_p(\vec{r},\vec{r^{\prime}})\right)^2\Bigg\}, \\ \mathcal{H}_{zr}&=&\frac{t_3}{4} \rho^{\alpha}(\vec{r})\left[(2+x_3)\rho^2(\vec{r})-(2x_3+1)(\rho_n^2(\vec{r})+\rho_p^2(\vec{r}))\right], \\ \mathcal{H}_{coul}&=&\frac{1}{2}\int d\vec{r^{\prime}}\frac{\rho_p(\vec{r})\rho_p(\vec{r^{\prime}})}{\mid\vec{r}-\vec{r^{\prime}}\mid}-\frac{3}{4}\left(\frac{3}{\pi}\right)^{1\over 3}\rho_p^{4\over 3}(\vec{r}), \\ \mathcal{H}_{SO}&=&-\frac{1}{2}W_0\left[\rho(\vec{r})\vec{\nabla} \cdot\vec{J}+\rho_n(\vec{r})\vec{\nabla}\cdot\vec{J_n}+\rho_p(\vec{r})\vec{\nabla}\cdot\vec{J_p}\right], \label{eq49}
\end{eqnarray}

respectively. The key quantity for writing the different contributions of the non-local HF energy density that appear in Eq.(\ref{eq43}) is the one-body density matrix, defined at HF level as

\begin{equation}
\sum_{i=1}^{A} \phi_i^*(\vec{r})\phi_i (\vec{r'})=\rho(\vec{R} + \frac{\vec{s}}{2},\vec{R} - \frac{\vec{s}}{2}) \label{eq50},
\end{equation}

where $\phi_i (\vec{r})$ are the single-particle wavefunctions, and $\vec{R}$ and $\vec{s}$ are the center of mass and relative coordinates, respectively.  From this density matrix the particle, kinetic energy and spin densities can be obtained as

\begin{equation}
\rho({\bf R}) = \rho(\vec{R},\vec{s})\vert_{s=0}, \label{eq51},
\end{equation}

\begin{equation}
\tau({\bf R})=\big(\frac{1}{4}\Delta_R - \Delta_s\big)\rho(\vec{R},\vec{s})\vert_{s=0}, \label{eq52}
\end{equation}

and

\begin{equation}
{\bf J}(\vec{R})= -i\bigg[{\bf \sigma} \times \big(\frac{1}{2}{\vec{\nabla_R}} +{\vec{\nabla_s}}\big)\bigg]\rho(\vec{R},\vec{s})\vert_{s=0} \label{eq53},
\end{equation}

respectively. \\

To obtain the quasi-local energy density functional one starts by replacing the HF density matrix in the energy density (\ref{eq43}) by the corresponding ETF counterpart reported in Refs.\cite{gridnev98,soubbotin00}. The ETF density matrix consists of a $\hbar^0$ (TF) part, which corresponds to the HF density matrix in the homogeneous system (Slater approach), plus a $\hbar^2$ contribution that takes into account the inhomogeneities of the system through second order derivatives of the proton and neutron densities. This $\hbar^2$ contribution also depends on the finite-range interaction by means of the inverse of the position and momentum dependent effective mass defined in Eqs.(\ref{eq18}) and (\ref{eq19}) and its momentum and position derivatives, where all these quantities are computed at the Fermi momentum \cite{gridnev98,soubbotin00}. Using this ETF density matrix in Eqs.(\ref{eq44}-\ref{eq49}) allows to write the HF energy at  ETF level. In particular, the ETF exchange energy can also be written as the sum of the $\hbar^0$ and the $\hbar^2$ contributions as 

\setlength\mathindent{0pt} 
\begin{eqnarray}
\mathcal{H}_{exch,0}&=&-\frac{1}{2}\sum_i\int d\vec{s} e^{-\frac{s^2}{\mu_i^2}}\{\left(B_i+\frac{W_i}{2}\right)\left[\left(\rho_n(\vec{r})\frac{3j_1(k_{F_n}s)}{k_{F_n}s}\right)^2+\left(\rho_p(\vec{r})\frac{3j_1(k_{F_p}s)}{k_{F_p}s}\right)^2\right]\nonumber \\ &&-\left(M_i+\frac{H_i}{2}\right)\rho_n(\vec{r})\frac{3j_1(k_{F_n}s)}{k_{F_n}s}\rho_p(\vec{r})\frac{3j_1(k_{F_p}s)}{k_{F_p}s}\} \label{eq54}
\end{eqnarray}

where $j_1$ is the $l=1$ spherical Bessel and

\begin{equation}
\mathcal{H}_{exch,2}=\sum_q\frac{\hbar^2}{2m_q}\bigg\{(f_q-1)\left(\tau_q-\frac{3}{5}k_{F_q}^2\rho_q-\frac{1}{4}\Delta\rho_q\right)+k_{F_q}f_{qk}\left[\frac{1}{27}\frac{(\vec{\nabla}\rho_q)^2}{\rho_q}-\frac{1}{36}\Delta\rho_q\right]\bigg\}, \nonumber \\ \label{eq55}
\end{equation}
\setlength\mathindent{20pt}

respectively. Notice that at ETF level the kinetic energy density in (\ref{eq44}) and (\ref{eq55}) is also the semi-classical one obtained using Eq.(\ref{eq52}) with the ETF density matrix. The explicit expression of this kinetic energy density, which is also a functional of the neutron and proton densities and their gradients, can be found in Refs.\cite{gridnev98,soubbotin00}. It is very important to emphasize that the resulting semiclassical ETF counterpart of the HF energy density becomes a functional of the neutron and proton densities, as it happens in the case of the zero-range Skyrme forces, as far as we have written the exchange energy in a local form. As it is well known, semi-classical quantities at ETF level are free from shell effects, which according to the Strutinsky energy theorem can be added perturbatively \cite{brack72}. However, as is pointed out in \cite{gridnev98}, and discussed in a more formal way in \cite{soubbotin03}, another possible way for recovering shell effects is to use the Kohn-Sham (KS) scheme \cite{kohn65} within the Density Functional Theory. To this end we write in the local ETF approach to the HF energy the particle, kinetic energy and spin densities for each type of particles ($q=n,p$) using a Salter determinant wave function built up with single particle orbitals $\phi_i$:

\begin{eqnarray}
\label{eq56}
\rho_q(\vec{r})=\sum_{i=1}^{A_q}\sum_{\sigma}\mid{\phi_i(\vec{r},\sigma,q)}\mid^2 \quad \tau_q(\vec{r})=\sum_{i=1}^{A_q}\sum_{\sigma}\mid{\vec{\nabla}\phi_i(\vec{r},	\sigma,q)}\mid^2, \nonumber \\ \vec{\mathbf{J}}_q(\vec{r})=i\sum_{i=1}^{A_q}\sum_{\sigma\sigma^{\prime}}\phi^*(\vec{r},\sigma,q) \left[(\vec{\sigma})_{\sigma\sigma^{\prime}}\times\vec{\nabla}\right]\phi_i(\vec{r},\sigma,q).
\end{eqnarray}

By applying now the variational principle to this local ETF energy density functional using the single particle orbitals $\phi_i$ as variational parameters one finds the following set of single particle equations: 

\begin{equation}\label{eq57}
{h}\phi_i=\left\{-\vec{\nabla}\frac{\hbar^2}{2m_q^*(\vec{r})}\vec{\nabla}+U_q(\vec{r})-i\vec{W}_q(\vec{r})(\vec{\nabla}\times\vec{\sigma}) \right\}\phi_i=\epsilon_i\phi_i
\end{equation}

where the effective mass, single-particle potential and form factor of the spin-orbit potential are given by:

\begin{eqnarray}\label{eq58}
\frac{\hbar^2}{2m_q^*}&=&\frac{\delta \mathcal{H}}{\delta \tau_q}, \\ U_q&=&\frac{\delta \mathcal{H}}{\delta \rho_q}, \\ \vec{W}_q&=&\frac{\delta \mathcal{H}}{\delta \vec{J}_q}.
\end{eqnarray}

The self-consistent solution of the set of Eqs.(\ref{eq57}) provides the single-particle orbitals that are used in the local ETF energy density functional through Eqs.(\ref{eq56}) to compute the HF energy in the quasilocal approximation, which allows to estimate the ground-state observables of spherical nuclei with a simplicity similar to that of similar calculations using Skyrme interactions. This self-consistent calculation also allows to obtain the Coulomb potential, including the exchange contribution at Slater level, as:

\begin{equation}
V_c(\vec{r}) =  \int d\vec{r^{\prime}}\frac{\rho_p(\vec{r^{\prime}})}{\mid\vec{r}-\vec{r^{\prime}}\mid}-\left(\frac{3}{\pi}\right)^{1\over 3}\rho_p^{1\over 3}(\vec{r}) \label{eq61},
\end{equation}

and the form factor of the spin-orbit potential for each type of particles as;

\begin{equation}
\vec{W}_q(\vec{r}) = \frac{1}{2} W_0 \big(\vec{\nabla}\rho(\vec{r}) +\vec{\nabla}\rho_q(\vec{r})\big),
\label{eq62}
\end{equation}  

from where the spin-orbit potential in (\ref{eq57}) and, consequently, in the real part of the optical potential can be written in the usual way as

\begin{equation}
V_{SO,q}(\vec{r}) = \frac{1}{r}\vec{W}_q(\vec{r}) \vec{l}\cdot\vec{s} \label{eq63}.
\end{equation}

The QEDFT introduced in Ref.\cite{soubbotin03} has been generalized to include pairing correlations in Ref.\cite{krewald06}. The agreement between our QEDFT estimates, which can include pairing correlations by means of an improved BCS calculation, is in tune with full HFB results as it can be seen in Refs.\cite{gridnev98,soubbotin03,krewald06,behera16}. These results support the usage of using the self-consistent quasi-local neutron and proton densities of target nuclei in the MOP calculation with Gogny forces. As an example of the QEDFT results, we display in Figure \ref{fig16} the neutron and proton densities of the nucleus $^{208}$Pb compared with a full HF calculation performed in a harmonic oscillator basis \cite{robledo02}. We see that the QEDFT densities reproduce nicely the HF profiles with slightly more pronounced oscillations.

\end{document}